\documentclass[a4paper, reprint, aps, prb, nofootinbib, groupedaddress]{revtex4-2} 

\usepackage{graphicx}
\usepackage{amsmath,amsthm,amssymb,amsfonts}
\usepackage{physics,mathrsfs,mathtools,array,bm}
\usepackage{newtxtext,newtxmath}
\usepackage{orcidlink}
\usepackage{hyperref}
\hypersetup{
  pdftitle  = {Revisiting magnetoelectric response in collinear antiferromagnetic zigzag chains: A downfolding approach beyond conventional low-energy models}, 
  pdfauthor = {Shuhei Kanda, Satoru Hayami}, 
  colorlinks=true,linkcolor=blue,urlcolor=blue,citecolor=blue
}

\newcommand{\kB}{k_{\mathrm{B}}}
\newcommand{\muB}{\mu_{\mathrm{B}}}
\newcommand{\eff}{\mathrm{eff}}

\begin{document}

\title{Revisiting magnetoelectric response in collinear antiferromagnetic zigzag chains:\\A downfolding approach beyond conventional low-energy models}
\author{Shuhei Kanda\,\orcidlink{0009-0007-0852-9979}}
\author{Satoru Hayami\,\orcidlink{0000-0001-9186-6958}}
\affiliation{Graduate School of Science, \href{https://ror.org/02e16g702}{Hokkaido University}, Sapporo 060-0810, Japan \vspace{4pt}} 
\date{\today}

\begin{abstract}
Magnetoelectric (ME) effects in antiferromagnets provide a fertile platform for exploring symmetry-driven cross-correlated responses. 
However, their microscopic origin remains elusive and is often obscured in simplified low-energy descriptions.
In this study, we revisit the microscopic mechanism of the ME effect in a collinear antiferromagnetic zigzag chain by employing a multi-orbital tight-binding model that explicitly includes both $s$- and $p$-orbital degrees of freedom. 
Using analytical and numerical calculations based on the Kubo formula, we demonstrate that the ME response is governed by orbital degrees of freedom activated through $s$--$p$ hybridization, while the spin contribution vanishes due to spin conservation. 
To elucidate the low-energy description, we derive an effective Hamiltonian projected onto the $s$-orbital subspace using the Schur complement. 
We show that a naive application of the Kubo formula within this effective model fails to capture the ME response. 
This issue is resolved by systematically incorporating vertex corrections in terms of orbital hybridization into the response functions. 
Furthermore, by introducing a quasiparticle renormalization scheme, we formulate a renormalized Kubo formula that preserves conservation laws and accurately reproduces the full multi-orbital results.
Our analysis revisits the conventional low-energy perspective and reveals that the ME effect originates from virtual interorbital processes encoded in vertex corrections, rather than from the bare low-energy Hamiltonian. 
The effective framework developed here provides a unified microscopic understanding of orbital-driven ME responses and offers a systematic route to incorporate hybridization effects beyond simple low-energy models.
\end{abstract}
\maketitle

\section{introduction} \label{sec:introduction}

Multiferroics, characterized by the simultaneous breaking of space-inversion and time-reversal symmetries, have long attracted considerable attention owing to their ability to host physical phenomena, 
such as cross-correlated responses including 
the magnetoelectric (ME) effect~\cite{Fiebig_2005,Spaldin_2008,KhomskiiPhysics.2.20, tokura2014multiferroics,doi:10.7566/JPSJ.87.033702,Hayami_PhysRevB.90.081115,PhysRevB.97.134423,Shitade_PhysRevB.98.020407,thole2018magnetoelectric,ding2021field,PhysRevB.105.155157,PhysRevB.111.L201112}.
Such multiferroic states are often characterized by a finite magnetic toroidal dipole (MTD) moment, also known as the toroidal moment, 
a time-reversal-odd polar vector~\cite{DUBOVIK1990145,Spaldin_2008,kopaev2009toroidal,hayami2016emergent,hayami2024unified}, 
which reflects an intrinsic asymmetry in the underlying electronic structure. 
As a consequence, a variety of unconventional properties emerge, including 
the nonlinear Hall effect~\cite{PhysRevLett.124.067203,PhysRevLett.127.277201,Hayami_PhysRevB.106.024405, PhysRevB.107.155109},
the nonreciprocal directional dichroism~\cite{PhysRevLett.95.237402,doi:10.1143/JPSJ.81.023712,Hayami_doi:10.7566/JPSJ.85.053705,PhysRevB.103.L180410,10.1063/5.0089235}, and
the nonreciprocal conduction~\cite{tokura2018nonreciprocal,doi:10.7566/JPSJ.91.115001,PhysRevB.105.155157,doi:10.7566/JPSJ.94.083705,13pd-tlzp}.
These phenomena have been widely discussed in systems such as spiral magnets, noncollinear antiferromagnets (AFMs), and low-dimensional magnets with local inversion asymmetry irrespective of the presence and absence of the spin--orbit coupling~\cite{Mostovoy_PhysRevLett.96.067601, tokura2014multiferroics, Hayami_PhysRevB.106.014420}, 
highlighting multiferroics as a central platform for exploring symmetry-driven novel electromagnetic responses.

A quasi-one-dimensional zigzag-chain system with staggered collinear AFM order provides a minimal platform for hosting MTD degrees of freedom.
Zigzag-chain AFMs are realized in several materials, including rare-earth-based compounds, where localized moments form a zigzag arrangement. 
In these systems, the AFM order is stabilized either by an Ising-type AFM exchange interaction~\cite{doi:10.7566/JPSJ.84.064717,Hayami_doi:10.7566/JPSJ.85.053705,PhysRevB.105.155157} or 
can be effectively induced by external control, such as an applied electric current~\cite{doi:10.7566/JPSJ.83.014703}.
Recent experiments have demonstrated nonreciprocal transport in NdRu$_{2}$Al$_{10}$, which hosts a collinear AFM zigzag-chain structure~\cite{13pd-tlzp}.
These results have renewed interest in AFM zigzag chains as a platform for investigating nonreciprocal transport and related symmetry-driven phenomena.

Despite significant progress in understanding nonreciprocal charge transport in collinear AFM zigzag chains, 
the ME response remains less explored from a theoretical perspective~\cite{cysne2021orbital, PhysRevB.105.155157}.
Previous studies based on a minimal tight-binding model that incorporates only the $s$-orbital degree of freedom have shown that this limited orbital description leads to rather restrictive conditions for the emergence of the ME effect, such as the requirement of a stacked structure~\cite{PhysRevB.105.155157}.
In other words, a minimal collinear AFM zigzag chain consisting of a single chain does not exhibit the ME effect within an $s$-orbital description, whereas real materials typically involve multiple orbital degrees of freedom and orbital hybridization that influences the low-energy electronic states.
This discrepancy suggests that such minimal models may be insufficient for capturing the ME effect in a single-chain system. 
Thus, it is crucial to establish a proper framework for deriving effective low-energy models starting from multi-orbital descriptions. 
For response phenomena, however, downfolding the Hamiltonian alone is generally insufficient, since the eliminated high-energy degrees of freedom can also modify the operators entering the response functions.
Clarifying how these effects are encoded in a low-energy theory is therefore essential for constructing a consistent description of the ME response.
In addition, the ME effect is of particular importance not only as an intriguing physical phenomenon, but also due to its close connection to the toroidization~\cite{PhysRevB.97.134423}, a gauge-invariant thermodynamic expression of the MTD moment.
These considerations highlight the need for a more general microscopic understanding of the ME response in zigzag-chain antiferromagnets.

In this paper, we revisit the microscopic mechanism of the ME effect in a collinear AFM zigzag chain described by a tight-binding model with $s$- and $p$-orbital degrees of freedom. 
By performing analytical and numerical calculations based on the Kubo formula, we demonstrate that the emergence of the ME effect is governed by orbital degrees of freedom.
We further construct a low-energy effective Hamiltonian projected onto the $s$-orbital subspace and clarify how the effects of orbital degrees of freedom can be incorporated into the effective model. 
In particular, we show that the ME response can be quantitatively captured within the $s$-orbital energy regime by incorporating vertex corrections and renormalization factors originating from $s$--$p$ hybridization.
We also discuss the accuracy of the approximation and analyze the decomposition of the contributions to the ME response.

The remainder of this paper is organized as follows.
In Sec.~\ref{sec:spmodel}, we present numerical results for the frequency dependence of the ME effect in a model including $s$- and $p$-orbital degrees of freedom.
In Sec.~\ref{sec:effHam}, we derive an effective Hamiltonian projected onto the $s$-orbital subspace.
In Sec.~\ref{sec:effSus}, we expand the Green's function using the Schur complement and discuss the corrections necessary to describe the ME response within the effective Hamiltonian.
In Sec.~\ref{sec:AC}, we introduce the analytic continuation procedure to obtain the real-frequency dependence of the ME response and present the Kubo formula with high-energy contributions.
In Sec.~\ref{sec:disc}, we discuss the ME effect within the framework of the effective Hamiltonian.
In Secs.~\ref{subsec:conpairapp}--\ref{subsec:contri}, we compare the results of the effective model with those of the $s$--$p$ model and decompose the contributions to the ME response.
In Sec.~\ref{subsec:selrule}, we analyze the selection rules to identify the essential model parameters.
Sec.~\ref{sec:Summary} summarizes this paper.
Appendices include the full expressions for the corrections in Appendix~\ref{sec:app:Operators},
verification of the conservation laws in the effective model in Appendix~\ref{sec:app:WTi}, and 
the details of the numerical analytic continuation in Appendix~\ref{sec:app:ACFlow}.

\section{Magnetoelectric response in the $s$--$p$ model} \label{sec:spmodel} 

\subsection{$s$-$p$ Model Hamiltonian}\label{subsec:spHam}

\begin{figure}[t]
  \centering
  \includegraphics[width=0.70\linewidth]{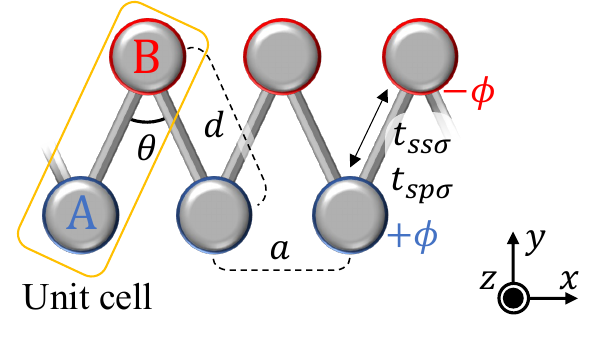}
  \caption{\label{fig:zigzag}
  Schematic picture of the AFM zigzag-chain $s$-$p$ model.
  Here, $\theta$ and $d$ denote the apex angle and the nearest-neighbor distance between A and B sublattices, respectively.
  The parameters $t_ {ss\sigma}$ and $t_{sp\sigma}$ represent the Slater-Koster hopping integrals~\cite{Slater_PhysRev.94.1498}.
  The parameter $\phi$ denotes the AFM molecular field.
  }
\end{figure}

We consider a minimal multi-orbital model consisting of $s$- and $p$-orbital degrees of freedom in a single zigzag chain along the $x$ direction, as shown in Fig.~\ref{fig:zigzag}.
The tight-binding Hamiltonian is given by
\begin{align}
  \mathcal{H}
  =\sum_{k}\sum_{\tau\tau'} \sum_{mm'\sigma\sigma'} 
  \left[\hat{h}(k)\right]^{\tau\tau'}_{ mm'\sigma\sigma'}\, c_{k\tau m\sigma}^{\dagger} \, c_{k\tau' m'\sigma'},
\end{align}
where $c_{k\tau m\sigma}$ and $c_{k\tau m\sigma}^{\dagger}$ are the fermionic annihilation and creation operators with the crystal momentum along the $x$ direction $\hbar k$, the sublattice $\tau=\mathrm{A},\mathrm{B}$, the $s$-$p$ orbital $m=s,p_x,p_y,p_z$, and the spin $\sigma=\uparrow,\downarrow$.
The Hamiltonian consists of four terms as follows:
\begin{align}\label{eq:spHam}
  \mathcal{H}= H_{\mathrm{hop}}+ H_{\Delta}+H_{\mathrm{MF}}-\mu N,
\end{align}
where $\mu$ and $N=\sum_{k}\sum_{\tau m\sigma} c_{k\tau m\sigma}^{\dagger}c_{k\tau m\sigma}$ are chemical potential and particle number operator, respectively.

The first term $H_{\mathrm{hop}}$ describes the electron hopping, incorporating the Slater--Koster parameters $t_{ss\sigma}$ and $t_{sp\sigma}$ for nearest-neighbor processes~\cite{Slater_PhysRev.94.1498}.
For simplicity, the hopping between the $p$ orbitals is neglected ($t_{pp\sigma}=t_{pp\pi}=0$).
The hopping Hamiltonian is given by 
\begin{align}\begin{aligned}\label{eq:hop}
  H_{\mathrm{hop}}
  =\sum_{k} \sum_{mm'\sigma}\left(
  \varepsilon_{mm'}(k) \, c_{k\mathrm{B} m\sigma}^{\dagger} c_{k\mathrm{A} m'\sigma}
  + \varepsilon_{mm'}^{*}(k) \, c_{k\mathrm{A} m'\sigma}^{\dagger} c_{k\mathrm{B} m\sigma}
  \right),
\end{aligned}\end{align}
where
\begin{subequations}\begin{align}
  \varepsilon_{ss}(k)   &= -2t_{ss\sigma}\cos\frac{ka}{2}e^{ika/2},\label{eq:sshop}\\
  \varepsilon_{sp_x}(k)  = -{\varepsilon}_{p_xs}(k)
                              &= 2it_{sp\sigma}\sin\frac{\theta}{2}\sin\frac{ka}{2}e^{ika/2},\\
  \varepsilon_{sp_y}(k)  = -{\varepsilon}_{p_ys}(k)
                              &= -2 t_{sp\sigma}\cos\frac{\theta}{2}\cos\frac{ka}{2}e^{ika/2},\\
  \varepsilon_{mm'}(k)  &=0\quad(\mathrm{otherwise}),
\end{align}\end{subequations}
with $a = 2d \sin (\theta/2)$.
Here, $\theta$ and $d$ denote the apex angle and the nearest-neighbor distance between A and B sublattices, respectively.
The second term $H_{\Delta}$ represents the energy offset of the $p$ orbital relative to the $s$ orbital,
which is expressed as
\begin{align}
  H_{\Delta}
  = \Delta_p \sum_{k}\sum_{\tau} \sum_{m\sigma}^{m=p_x,p_y,p_z}
  \, c_{k\tau m\sigma}^{\dagger} \, c_{k\tau m\sigma}.
\end{align}

The third term $H_{\mathrm{MF}}$ describes the molecular field associated with the collinear AFM order with the $z$-directional polarization,
which is given by
\begin{align}
  H_{\mathrm{AF}} &= \frac{\phi}{\hbar} (L^{\mathrm{AF}}+\mathrm{g}S^{\mathrm{AF}}),\\
  L^{\mathrm{AF}} &= \sum_{k} \sum_{mm'\sigma} l^{z}_{mm'} 
  \,(c_{k\mathrm{A} m\sigma}^{\dagger} \, c_{k\mathrm{A} m'\sigma} - c_{k\mathrm{B} m\sigma}^{\dagger} \, c_{k\mathrm{B} m'\sigma}),\\
  S^{\mathrm{AF}} &= \sum_{k} \sum_{m\sigma\sigma'} s^{z}_{\sigma\sigma'} 
  \,(c_{k\mathrm{A} m\sigma}^{\dagger} \, c_{k\mathrm{A} m\sigma'} - c_{k\mathrm{B} m\sigma}^{\dagger} \, c_{k\mathrm{B} m\sigma'}),
\end{align}
where $\bm{l}_{mm'}$ and $\bm{s}_{\sigma\sigma'}$ denote the matrix elements of the orbital and spin angular momentum operators, respectively.
$\hbar$ is the reduced Planck constant and $\mathrm{g}$ is the Land\'e $\mathrm{g}$ factor.
The AFM molecular field can be realized either 
through a Hartree-type mean-field approximation of the Ising-type AFM exchange interaction~\cite{doi:10.7566/JPSJ.84.064717,Hayami_doi:10.7566/JPSJ.85.053705,PhysRevB.105.155157} or 
through external control achieved by applying an electric current~\cite{doi:10.7566/JPSJ.83.014703}.
This AFM spin configuration corresponds to a state with a finite MTD moment along the $x$ direction~\cite{doi:10.7566/JPSJ.83.014703, doi:10.7566/JPSJ.84.064717}.

In this study, atomic spin--orbit coupling in $p$-orbital sector is not introduced explicitly; 
instead, the AFM molecular field arising from orbital magnetization plays a crucial role. 
In realistic situations, however, this molecular field is typically mediated by spin--orbit coupling.

\subsection{Kubo formula}\label{subsec:spKubo}

\begin{figure*}[t]
  \centering
  \includegraphics[width=0.85\linewidth]{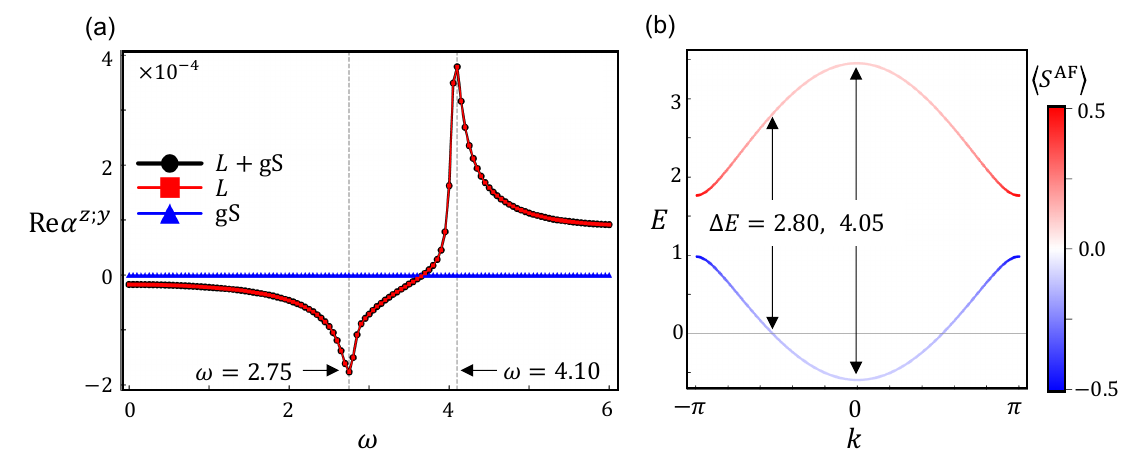}
  \caption{\label{fig:spres}
  (a) 
  Frequency $\omega$ dependence of the real part of the ME tensor component $\alpha^{z;y}(\omega)$ for contributions from the atomic orbital ($L$) and spin ($\mathrm{g}S$) magnetizations.
  The spin contribution vanishes, while only the orbital contribution remains finite.
  Peaks are observed at $\omega=2.75$ and $4.10$, corresponding to the minimum and maximum interband transition energies in panel (b).
  (b) 
  Band structure of the $s$--$p$ model Hamiltonian, where the electronic states are primarily characterized by the $s$ orbital due to the large energy separation $\Delta_{p}$. 
  The color bar represents the expectation value of the AFM spin component with $k$ and the band index.
  The horizontal line indicates the Fermi level.
  In both (a) and (b), the parameters are set to $t_{ss\sigma}=1, t_{sp\sigma}=0.7, \Delta_{p}=10, \phi=0.4, \theta={3\pi}/{5},\mu=-1.5$, $N_k=2^7$, $T=0.01$, and $\gamma=0.05$. 
  }
\end{figure*}

We evaluate the ME tensor for the above model within linear response theory.
The ME tensor $\alpha^{\nu;\rho}$ is defined as the linear-response
coefficient relating the induced magnetization $\delta M^{\nu}$ to an external electric field $E^{\rho}$:
\begin{align}
  \delta M^{\nu} (\omega) = \sum_{\rho} \alpha^{\nu;\rho} (\omega) E^{\rho}(\omega),
\end{align}
where $\omega$ is the frequency.
In the presence of the AFM molecular field, a finite MTD moment is induced along the $x$ direction, which allows
a finite $\alpha^{z;y}$ component of the ME tensor by symmetry~\cite{PhysRevB.98.165110, PhysRevB.98.245129, PhysRevB.104.054412, PhysRevB.105.155157, hayami2024unified}.

The linear ME tensor is calculated using the Kubo formula:
\begin{align}\begin{aligned}\label{eq:KuboFml}
  \alpha^{\nu;\rho}(\omega) & \\
  =-\frac{i\hbar}{V} & \sum_{k}\sum_{\alpha\beta}\frac{f(E_{k\alpha})-f(E_{k\beta})}{E_{k\alpha}-E_{k\beta}}
  \frac{  M^{\nu}_{k;\alpha\beta} J^{\rho}_{k;\beta\alpha}}{E_{k\alpha}-E_{k\beta}+\hbar\omega+i\hbar\gamma},
\end{aligned}\end{align}
where 
$E_{k\alpha}$ are the eigenvalues of the Bloch Hamiltonian matrix $\hat{h}(k)$,
$f(E)=(e^{E/\kB T}+1)^{-1}$ is the Fermi distribution function, 
$V=2N_k d^2$ is the volume of the system, $N_k$ is the number of $k$-mesh points, 
and $\gamma$ is the scattering rate, characterized as the inverse of the relaxation time under the relaxation time approximation.

The matrix elements $M^{\nu}_{k;\alpha\beta}$ are obtained from the magnetization operator, which is given by
\begin{align}
  M^{\nu} 
  &= \frac{\muB }{\hbar} \left( L^{\nu} + \mathrm{g} S^{\nu} \right),\\
  L^{\nu}
  &= \sum_{k}\sum_{\tau}\sum_{mm'\sigma} l^{\nu}_{mm'}\,c_{k\tau m\sigma}^{\dagger} \, c_{k\tau m'\sigma},\\
  S^{\nu}
  &= \sum_{k}\sum_{\tau}\sum_{m\sigma\sigma'} s^{\nu}_{\sigma\sigma'}\,c_{k\tau m\sigma}^{\dagger} \, c_{k\tau m\sigma'},
\end{align}
where $L^{\nu}$ and $S^{\nu}$ denote the orbital and spin angular momentum operators, respectively.
We neglect the contribution from the modern theory of orbital magnetization~\cite{PhysRevLett.102.146805,PhysRevB.83.085108,arXiv:2512.05862,arXiv:2512.19380}.

The matrix elements $J^{\nu}_{k;\alpha\beta}$ are obtained from the current operator, which is given by
\begin{subequations}\begin{align}
  J^{\nu} 
  = \sum_{k}\sum_{\tau\tau'} \sum_{mm'\sigma\sigma'} 
  &\left[\frac{e}{\hbar}  \frac{D\hat{h}(k)}{D k_{\nu}}\right]^{\tau\tau'}_{ mm'\sigma\sigma'}\, c_{k\tau m\sigma}^{\dagger} \, c_{k\tau' m'\sigma'},\\
  \frac{D\hat{h}(k)}{D k_{\nu}} \label{eq:Jmatrix}
  &= \Big.\frac{\partial\hat{h}(k_x)}{\partial k_{\nu}}\Big|_{k_x=k}+\frac{1}{i}\left[\hat{d}_{\mathrm{i}}^{\nu},\hat{h}(k)\right],
\end{align}\end{subequations}
where the first and second terms in the second line correspond to the intercell and intracell contributions to the velocity operator~\cite{PhysRevB.108.085403}.
Here, $e$ is the elementary charge and the square bracket $[\sharp_1,\sharp_2]=\sharp_1\sharp_2-\sharp_2\sharp_1$ is the commutator.
The intracell position operator is given by $\hat{\bm{d}}_{\mathrm{i}}=-\bm{d}_{\mathrm{AB}}\hat{\tau}_3/2$, 
where $\hat{\tau}_3$ is the $z$ component of the Pauli matrices in sublattice space and $\bm{d}_{\mathrm{AB}}=d(\sin{{\theta}/{2}},\cos{{\theta}/{2}})$ is the position difference in unit cell.
In multi-orbital tight-binding models, additional contributions to the velocity operator may arise from interorbital polarization~\cite{gmnv-cwvr}; however, we neglect such contributions in the present study.

\subsection{Numerical results} \label{subsec:spNR}

We evaluate the frequency $\omega$ dependence of the ME tensor $\alpha^{z;y}$ for the AFM zigzag-chain $s$--$p$ model in Eq.~(\ref{eq:spHam}).
The model parameters are set to
\begin{align}\label{eq:HamPara}
  t_{ss\sigma}=1, \, t_{sp\sigma}=0.7, \, \Delta_{p}=10, \, \phi=0.4,\, \theta=\frac{3\pi}{5},
\end{align}
with the chemical potential $\mu=-1.5$ in the $s$-orbital energy range.
We also set $e=\hbar=\muB=d=1$, $\mathrm{g}=2$, the number of $k$-mesh point $N_k=2^7$, the temperature $T=0.01$, and the scattering rate $\gamma=0.05$. 
We focus on the frequency range $0 \leq \omega \leq 6$, which is dominated by the $s$-orbital states.

Figure~\ref{fig:spres}(a) shows the $\omega$ dependence of the real part of the ME tensor, decomposed into
the contributions from the atomic orbital ($L$) and spin ($\mathrm{g}S$) magnetizations.
The spin contribution vanishes, while only the atomic orbital contribution remains finite. 
The absence of the spin contribution is owing to spin conservation in the present model, i.e., $[\hat{h},\hat{s}^z]=0$, 
which prohibits spin-flip processes induced by the electric field.

Clear peak structures are observed at $\omega=2.75$ and $4.10$. 
These peaks originate from interband transitions between AFM spin-split bands. 
To demonstrate this, we show the corresponding band structure in Fig.~\ref{fig:spres}(b), where the color scale represents the expectation value of the AFM spin component as a function of $k$ and the band index.
One finds that the minimum and maximum interband transition energies, $\varDelta E = 2.80$ and $4.05$, respectively, are in good agreement with the peak positions in Fig.~\ref{fig:spres}(a).
The slight shifts of the peak positions can be understood from the structure of the Kubo formula.
The response function takes a Lorentzian-derivative form proportional to $x/(x^2+\gamma^2)$, whose extrema occur at $x=\pm\gamma$.
As a result, the peak positions are shifted by $ \pm \gamma $ from the band transition energies, leading to $\omega = 2.80 - 0.05$ and $\omega = 4.05 + 0.05$.

\section{Effective Model} \label{sec:effHam}

\begin{figure}[b]
  \centering
  \includegraphics[width=0.70\linewidth]{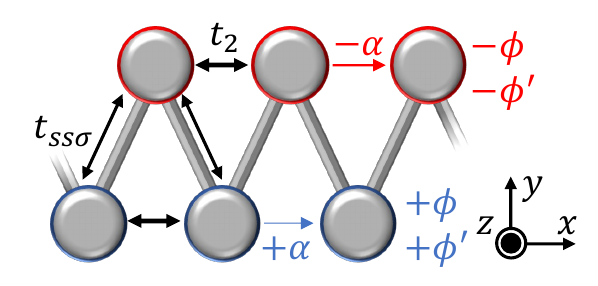}
  \caption{\label{fig:zigzageff}
  Schematic picture of the effective AFM zigzag-chain model.
  The parameters $t_2$, $\alpha$, and $\phi'$ stand for the effective next-nearest-neighbor hopping, antisymmetric spin--orbit interaction, and the effective molecular field, respectively.
  }
\end{figure}

In the following, we reformulate the ME response obtained for the $s$--$p$ model in terms of an effective model projected onto the $s$-orbital subspace.
For the model in Eq.~(\ref{eq:spHam}), the total Hilbert space $\mathbb{H}$ is decomposed as 
\begin{subequations}\begin{align}
  \mathbb{H}             &= \mathbb{H}_{\parallel} \oplus \mathbb{H}_{\perp}, \\
  \mathbb{H}_{\parallel} &= \mathrm{BZ}\otimes\{\mathrm{A,B}\}\otimes\{s\}\otimes\{\uparrow,\downarrow\}, \\
  \mathbb{H}_{\perp}     &= \mathrm{BZ}\otimes\{\mathrm{A,B}\}\otimes\{p_x,p_y,p_z\}\otimes\{\uparrow,\downarrow\},
\end{align}\end{subequations}
where $\mathbb{H}_{\parallel}$ and $\mathbb{H}_{\perp}$ denote the $s$-orbital and $p$-orbital subspaces, respectively.
The former corresponds to the low-energy subspace spanned by the $s$ orbitals, while the latter represents the higher-energy subspace composed of the $p$ orbitals.
Both subspaces include the wave vector, sublattice, and spin degrees of freedom in addition to the orbital degree of freedom, and BZ stands for the Brillouin zone. 
When we introduce projection operators $\hat{P}_{\parallel}$ and $\hat{Q}_{\perp}$ onto these subspaces, the Bloch Hamiltonian is written in block form as
\begin{align}
  \hat{h}(k) 
  = \begin{pmatrix}
    {\hat{h}_{\parallel}} & {\hat{\eta}^{\dagger}} \\ {\hat{\eta}} & {\hat{h}_{\perp}}
  \end{pmatrix} 
  = \begin{pmatrix}
    {\hat{P}_{\parallel} \hat{h}(k) \hat{P}_{\parallel}} & {\hat{P}_{\parallel} \hat{h}(k) \hat{Q}_{\perp}} \\ 
    {\hat{Q}_{\perp} \hat{h}(k) \hat{P}_{\parallel}} & {\hat{Q}_{\perp} \hat{h}(k) \hat{Q}_{\perp}}
  \end{pmatrix}.
\end{align}
Based on this decomposition, we derive the effective Hamiltonian projected onto the $s$-orbital subspace~\cite{10.1063/1.1724312,RevModPhys.36.1076,LEINAAS197819,10.1063/1.4904200}:
\begin{align}
  \mathcal{H}_{\eff}(z)
  &= \sum_{k}\sum_{\tau\tau'} \sum_{\sigma\sigma'} 
  \left[\hat{h}_{ \eff }(k,z)\right]_{\sigma\sigma'}^{\tau\tau'}\, c_{k\tau\sigma}^{\dagger} \, c_{k\tau'\sigma'}\\
  \hat{h}_{ \eff }(k,z) \label{eq:effhamdef}
  &=  \hat{h}_{\parallel} + \hat{\eta}^{\dagger} \, \frac{1}{\hbar z - \hat{h}_{\perp}} \, \hat{\eta},
\end{align}
where we omit the subscript for the $s$ orbital as $c_{k\tau\sigma}=c_{k\tau s\sigma}$.
This expression corresponds to the projected Schr\"{o}dinger equation as well as the Schur-complement form obtained by integrating out the $p$-orbital degrees of freedom.

\begin{figure*}[t]
  \centering
  \includegraphics[width=1.0\linewidth]{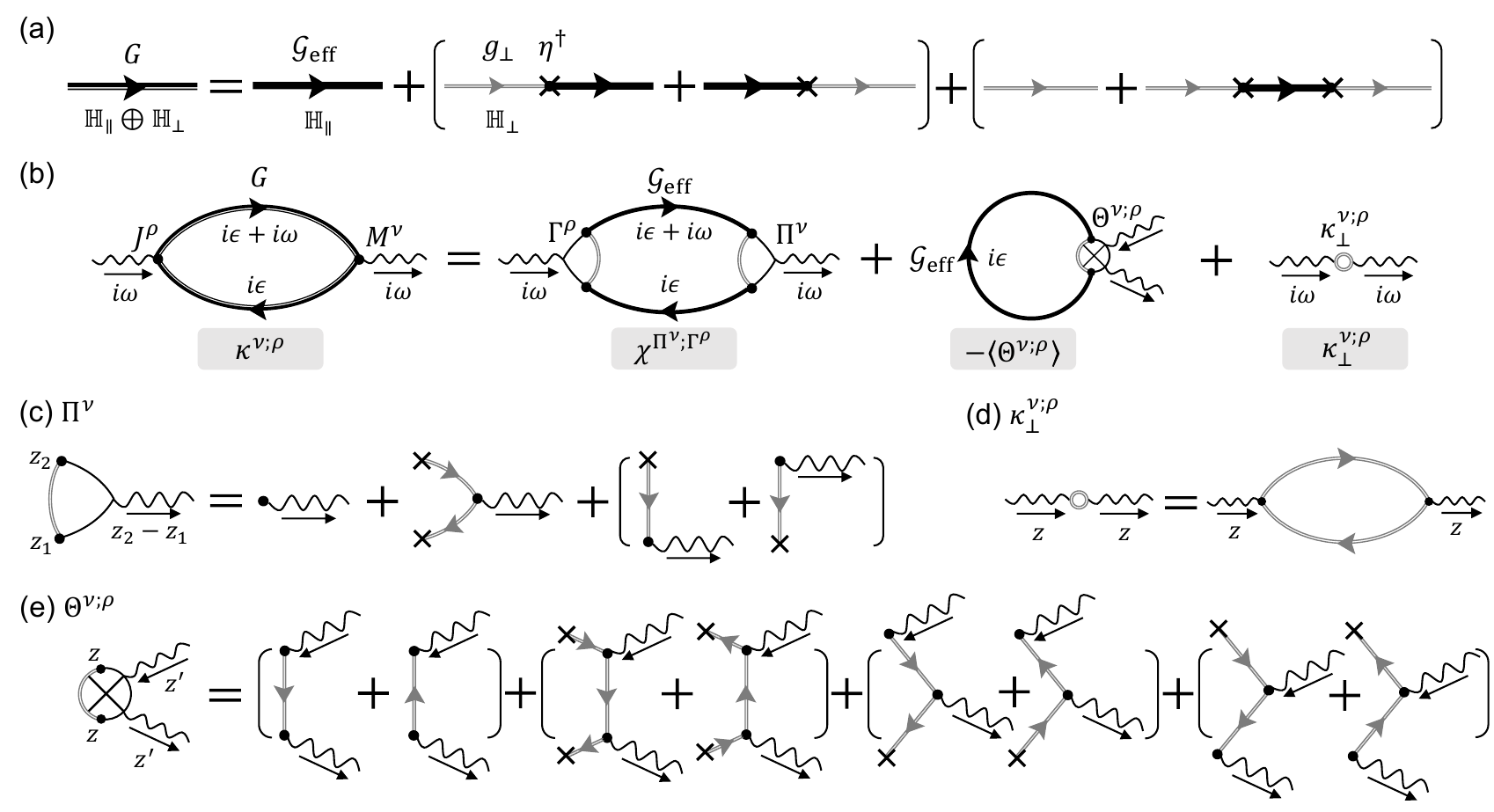}
  \caption{\label{fig:effsusdiag}
  (a) 
  Diagrams corresponding to the Schur complement for the Green's function in Eq.~(\ref{eq:schur}).
  The black double line, black thick line, and gray double line denote the Green's functions $\hat{G}$, $\hat{\cal{G}}_{\eff}$, and $\hat{g}_{\perp}$, 
  corresponding to the $s$--$p$ Hamiltonian, the effective Hamiltonian, and the Hamiltonian in the $p$-orbital subspace, respectively.
  The cross symbol ($\cross$) indicates the $s$--$p$ hybridization $\hat{\eta}$.
  (b) 
  Diagrams corresponding to the expanding ME response function $\kappa^{\nu;\rho}$ in Eq.~(\ref{eq:effsus}).
  On the left-hand side, the vertex point accompanied by a wavy line with outgoing (incoming) arrow represents the magnetization (electric current) operator $\hat{M}^{\nu}$ ($\hat{J}^{\rho}$).
  On the right-hand side, vertex corrections are included at the endpoints, resulting in dressed vertices, as illustrated in (c)--(e).
  (c)
  Diagrams representing the vertex-corrected effective magnetization operator $\hat{\Pi}^{\nu}$ in Eq.~(\ref{eq:effchi}).
  Reversing the direction of the arrow attached to the wavy line represents the effective current operator $\hat{\Gamma}^{\rho}$.
  The detailed expressions are given in Appendix~\ref{sec:app:Operators}.
  (d)
  Diagrams representing the ME response function defined entirely within the $p$-orbital subspace $\kappa^{\nu;\rho}_{\perp}$ in Eq.~(\ref{eq:effKperp}).
  This corresponds to a bubble diagram closed by $\hat{g}_{\perp}$.
  (e)
  Diagrams representing the magnetoelectric operator generated by the incorporation of $s$--$p$ hybridization $\hat{\Theta}^{\nu;\rho}$ in Eq.~(\ref{eq:effTheta}).
  }
\end{figure*}

The effective Hamiltonian is decomposed as
\begin{align}\label{eq:effham}
  \mathcal{H}_{ \eff }(z) = H'_{\mathrm{hop}} + H_{\mathrm{ASOI}} + H'_{\mathrm{MF}} - \mu N',
\end{align}
where $N'$ is the particle number operator in the $s$-orbital subspace.
A schematic illustration of this effective model is shown in Fig.~\ref{fig:zigzageff}.

The first term describes an effective hopping $H'_{\mathrm{hop}}$, which is represented as follows:
\begin{subequations}\begin{align}
  H'_{\mathrm{hop}}              &= H^{\mathrm{AB}}_{\mathrm{hop}} + H^{(2)}_{\mathrm{hop}},\\
  H^{\mathrm{AB}}_{\mathrm{hop}} &= \sum_{k} \sum_{\sigma} \left(
    \varepsilon_{ss}(k) c_{k\mathrm{B}\sigma}^{\dagger}c_{k\mathrm{A}\sigma} + \varepsilon_{ss}^{*}(k) c_{k\mathrm{A}\sigma}^{\dagger}c_{k\mathrm{B}\sigma}
  \right),\\
  H^{(2)}_{\mathrm{hop}}         &= \sum_{k} \sum_{\tau} \sum_{\sigma} \varepsilon(k,z) \, c_{k\tau\sigma}^{\dagger} c_{k\tau\sigma},
\end{align}\end{subequations}
where $\varepsilon_{ss}(k)$ defined as Eq.~(\ref{eq:sshop}), and 
\begin{align}
  \varepsilon(k,z) = -2 t_2(z) \left(1+\cos\theta\cos ka\right).
\end{align}
$t_2(z)$ denotes an effective next-nearest-neighbor hopping generated by virtual processes involving the $p$ orbitals, and for $\mathrm{g}=2$ it can be written as follows:
\begin{align}
  t_2(z) = 
  \frac{\left(\tilde{\Delta}^2-2\phi^2\right)t_{sp\sigma}^2}{\left(\tilde{\Delta}^2-4\phi^2\right)\tilde{\Delta}},
\end{align}
where $\tilde{\Delta}=\Delta_p-(\hbar z+\mu)$.

The second term describes an effective antisymmetric spin--orbit interaction (ASOI) $H_{\mathrm{ASOI}}$,
which is given by
\begin{align}\begin{aligned}
  &H_{\mathrm{ASOI}} \\ & \, = 
  \frac{2\alpha(z)}{\hbar} \sum_{k} \sum_{\sigma} \bm{g}(k)\cdot\bm{s}_{\sigma\sigma'} \,
  ( c_{k\mathrm{A}\sigma}^{\dagger}c_{k\mathrm{A}\sigma'} - c_{k\mathrm{B}\sigma}^{\dagger}c_{k\mathrm{B}\sigma'} ) ,
\end{aligned}\end{align}
where $\bm{g}(k)=(\sin\theta\sin ka)\vb{e}_z$,
\begin{align}
  \alpha(z) = -\frac{4\phi^2t_{sp\sigma}^2}{\left(\tilde{\Delta}^2-4\phi^2\right)\tilde{\Delta}
  }.
\end{align}
This term originates from virtual hopping processes involving the $p$ orbitals and acts as an effective ASOI in the $s$-orbital subspace.

The third term corresponds to an effective molecular field $H'_{\mathrm{MF}}$, which is represented as follows:
\begin{align}
  H'_{\mathrm{MF}} &= \frac{\phi}{\hbar} \mathrm{g} S'^{\mathrm{AF}} + \frac{\phi'_{\mathrm{o}}(z)}{\hbar} L^{\mathrm{AF}}_{ \eff } +  \frac{\phi'_{\mathrm{s}}(z)}{\hbar} \mathrm{g} S^{\mathrm{AF}}_{ \eff } , 
\end{align}
where $S'^{\mathrm{AF}}$ represents $S^{\mathrm{AF}}$ restricted to the $s$-orbital subspace.
The latter two terms represent corrections induced by the eliminated $p$-orbital degrees of freedom.
The effective operators are given by
\begin{subequations}\begin{align}
  L^{\mathrm{AF}}_{ \eff } &= \sum_{k} \sum_{\tau} \sum_{\sigma} l^{\mathrm{AF}}(k) \, c_{k\tau\sigma}^{\dagger} c_{k\tau\sigma} \\
  S^{\mathrm{AF}}_{ \eff } &= \sum_{k} \sum_{\sigma} \mathrm{g}'_{k} s^{z}_{\sigma\sigma'} (c_{k\mathrm{A}\sigma}^{\dagger} c_{k\mathrm{A}\sigma'} - c_{k\mathrm{B}\sigma}^{\dagger} c_{k\mathrm{B}\sigma'}),\\
  l^{\mathrm{AF}}(k) &= \hbar\sin\theta\sin ka,\quad \mathrm{g}'_{k}=1+ \cos\theta\cos ka,
\end{align}\end{subequations}
and 
\begin{align}
  \phi'_{\mathrm{o}}(z) &= \phi'_{\mathrm{s}}(z) = -\frac{2t_{sp\sigma}^2}{\tilde{\Delta}^2-4\phi^2}\phi .
\end{align}
These terms describe renormalized molecular fields arising from virtual processes involving the eliminated $p$-orbital degrees of freedom.

This effective Hamiltonian is qualitatively equivalent to that used in previous studies~\cite{doi:10.7566/JPSJ.83.014703,PhysRevB.105.155157}, except for the additional correction terms in the molecular-field sector.
To clarify the ME response within this effective model, let us consider a straightforward application of the Kubo formula to $\hat{h}_{\eff}(0)$. 
In this case, the spin contribution vanishes because the effective Hamiltonian preserves the $z$-component of spin, $[\hat{h}_{\eff}, \hat{s}^z]=0$, as discussed in Sec.~\ref{subsec:spNR}. 
On the other hand, the orbital contribution is also absent, since the atomic orbital angular momentum is not defined within the $s$-orbital subspace, i.e., $\hat{P}_{\parallel}\hat{l}^z\hat{P}_{\parallel}=0$. 
As a result, both spin and orbital contributions to the ME response vanish, and the ME effect cannot be captured within a naive Kubo formulation based solely on the effective $s$-orbital Hamiltonian. 
This conclusion is consistent with previous studies, where no finite ME response is obtained in similar effective models.
This issue can be resolved by incorporating corrections to the current and magnetization operators in the ME response tensor, as shown in the next section, which restores a finite ME response and reproduces the behavior of the original $s$--$p$ model.

\section{Expression of ME tensor in the effective model} \label{sec:effSus}

The ME tensor $\alpha^{\nu;\rho}$ can be formulated in terms of the Matsubara Green's function $\hat{G}$ through the ME response function $\kappa^{\nu;\rho}$ as follows~\cite{Altland_Simons_2010,PhysRevB.110.165111}:
\begin{align}\label{eq:spsusG}
  \alpha^{\nu;\rho}(\omega)    &= \frac{\left.\kappa^{\nu;\rho}(i\omega_m)\right|_{i\omega_m\rightarrow\omega+i\gamma}}{i(\omega+i\gamma)} \\
  \label{eq:spsusG2}
  \kappa^{\nu;\rho}(i\omega_m) &= -\frac{\kB T}{V} \sum_{\epsilon_n} \Tr \left[ \hat{M}^{\nu} \hat{G}(i\epsilon_n + i\omega_m) \hat{J}^\rho \hat{G}(i\epsilon_n) \right],
\end{align}
where $\epsilon_n=(2n+1)\pi\kB T/\hbar$ and $\omega_m=2m\pi\kB T/\hbar$ are the fermionic and bosonic Matsubara frequencies, respectively.
The trace ($\Tr$) includes $k$-summation.

Since the present model is noninteracting, the Matsubara Green's function is given by the resolvent,
$\hat{G}(k,i\epsilon_n) = \big[i\hbar\epsilon_n-\hat{h}(k)\big]^{-1}$.
Applying the Schur complement, the Green's function can be decomposed into contributions from the $s$-orbital subspace and its complement as
\begin{align}\label{eq:schur}
  \hat{G}(k,z)=\begin{pmatrix}
    {\hat{\mathcal{G}}_{\eff}} & {\hat{\mathcal{G}}_{\eff} \, \hat{\eta}^{\dagger} \, \hat{g}_{\perp}}\\
    { \hat{g}_{\perp} \, \hat{\eta} \, \hat{\mathcal{G}}_{\eff}} & { \hat{g}_{\perp} + \hat{g}_{\perp} \, \hat{\eta} \, \hat{\mathcal{G}}_{\eff} \hat{\eta}^{\dagger} \hat{g}_{\perp}}
  \end{pmatrix},
\end{align}
where $\hat{\mathcal{G}}_{\eff}(k, z)= \big[\hbar z - \hat{h}_{\eff}(k,z)\big]^{-1}$ and $\hat{g}_{\perp}(z)=\big[\hbar z - \hat{h}_{\perp}\big]^{-1}$.
This decomposition explicitly separates low-energy ($s$-orbital) and high-energy ($p$-orbital) processes.
Substituting Eq.~(\ref{eq:schur}) into Eq.~(\ref{eq:spsusG2}), the ME response function is decomposed as
\begin{equation}
  \begin{aligned}\label{eq:effsus}
    \kappa^{\nu;\rho}(i\omega_m) 
    &= \chi^{\Pi^\nu;\Gamma^\rho}(i\omega_m) - {\langle\Theta^{\nu;\rho}\rangle} + \kappa_{\perp}^{\nu;\rho}(i\omega_m).
  \end{aligned}
\end{equation}
Figures~\ref{fig:effsusdiag}(a) and (b) illustrate the diagrams corresponding to Eq.~(\ref{eq:schur}) and Eq.~(\ref{eq:effsus}), respectively.

The first term $\chi^{\Pi^\nu;\Gamma^\rho}(i\omega_m)$ describes the magnetization-current correlation function within the $s$-orbital subspace, 
including vertex corrections arising from $s$--$p$ hybridization:
\begin{equation}\begin{aligned}\label{eq:effchi}
  &\chi^{\Pi^\nu;\Gamma^\rho}(i\omega_m)\\
  &\,\,= -\frac{\kB T}{V}\sum_{\epsilon_n} \Tr_{\parallel} \left[ 
     \hat{\Pi}^\nu \hat{\mathcal{G}}_{\eff}(i\epsilon_n + i\omega_m) \hat{\Gamma}^\rho \hat{\mathcal{G}}_{\eff}(i\epsilon_n)
  \right],
\end{aligned}\end{equation}
where $\Tr_{\parallel}$ denotes the partial trace in $s$-orbital subspace $\mathbb{H}_{\parallel}$.
$\hat{\Pi}^{\nu}=\hat{\Pi}^{\nu}(i\epsilon_n,i\epsilon_{n+m})$ and $\hat{\Gamma}^{\rho}=\hat{\Gamma}^{\rho}(i\epsilon_{n+m},i\epsilon_n)$ are the effective magnetization operator and effective electric current operator described as
\begin{align}
  \label{eq:effMag}\hat{\Pi}^{\nu}(z_1,z_2)     &= \frac{\muB}{\hbar}\left(\mathrm{g}\hat{s}^{\nu}+\varDelta\hat{l}^{\nu}(z_1,z_2)+\mathrm{g}\varDelta\hat{s}^{\nu}(z_1,z_2)\right),\\
  \label{eq:effECu}\hat{\Gamma}^{\rho}(z_1,z_2) &= \hat{J}^{\rho}_{\parallel}+\varDelta\hat{J}^{\rho}(z_1,z_2).
\end{align}
The corrected terms are expressed as Fig.~\ref{fig:effsusdiag}(c). 
In our model in Eq.~(\ref{eq:spHam}), they are given by
\begin{subequations}\begin{align}
  \varDelta\hat{s}^{\nu}(z_1,z_2)  &= \hat{\eta}^{\dagger} \hat{g}_{\perp}(z_1) \, \hat{s}^{\nu}_{\perp} \, \hat{g}_{\perp}(z_2) \hat{\eta},\\
  \varDelta\hat{l}^{\nu}(z_1,z_2)  &= \hat{\eta}^{\dagger} \hat{g}_{\perp}(z_1) \, \hat{l}^{\nu}_{\perp} \, \hat{g}_{\perp}(z_2) \hat{\eta},\\
  \varDelta\hat{J}^{\rho}(z_1,z_2) &= \hat{\eta}^{\dagger} \hat{g}_{\perp}(z_1) \hat{j}^{\rho} + \hat{j}^{\rho\dagger} \hat{g}_{\perp}(z_2) \hat{\eta},
\end{align}\end{subequations}
where $\hat{s}^{\nu}_{\perp}=\hat{Q}_{\perp}\hat{s}^{\nu}\hat{Q}_{\perp}$, $\hat{l}^{\nu}_{\perp}=\hat{Q}_{\perp}\hat{l}^{\nu}\hat{Q}_{\perp}$ and $\hat{j}^{\rho}=\hat{Q}_{\perp}\hat{J}^{\rho}\hat{P}_{\parallel}$.
The full expressions in terms of the matrix elements of the respective subspaces are given in Appendix~\ref{sec:app:Operators}.
These corrections arise from virtual processes involving the eliminated $p$-orbital degrees of freedom and encode the missing orbital response not captured by the effective Hamiltonian.
Specifically, while the effective Hamiltonian intrinsically captures only the even-order hybridization processes, the vertex corrections are essential to account for the missing odd-order contributions.

The second term $\langle\Theta^{\nu;\rho}\rangle$ is a magnetoelectric operator term generated by the incorporation of $s$-$p$ hybridization defined by
\begin{align}\label{eq:effTheta}
  \langle\Theta^{\nu;\rho}\rangle(i\omega_m)
  = \frac{\kB T}{V}\sum_{\epsilon_n} \Tr_{\parallel} \left[ 
    \hat{\mathcal{G}}_{\eff}(i\epsilon_n) \hat{\Theta}^{\nu;\rho}
  \right],
\end{align}
where $\hat{\Theta}^{\nu;\rho}=\hat{\Theta}^{\nu;\rho}(i\epsilon_n;i\omega_m)$ is an effective magnetoelectric operator defined as Fig.~\ref{fig:effsusdiag}(e).
In our situation, it is given by
\begin{align}\begin{aligned}\label{eq:effThetaope}
  \hat{\Theta}^{\nu;\rho}(z;z')
  &       = \hat{\eta}^{\dag} \hat{g}_{\perp}(z) \hat{M}^{\nu}_{\perp} \hat{g}_{\perp}(z+z') \hat{j}^{\rho} \\
  &\quad\,+ \hat{j}^{\rho\dag}  \hat{g}_{\perp}(z-z') \hat{M}^{\nu}_{\perp} \hat{g}_{\perp}(z) \hat{\eta}.
\end{aligned}\end{align}
The effective magnetoelectric operator induced by hybridization can be expressed in terms of the commutator of the polarization and magnetization operators:
\begin{align}
  \hat{\Theta}^{\nu;\rho} \sim \frac{1}{i\hbar} [\hat{P}^{\nu},\hat{M}^{\rho}].
\end{align}
This term usually vanishes ($[\hat{P}^{\nu},\hat{M}^{\rho}]=0$), but it can be interpreted as a vertex-correction contribution within the projected subspace.
From a structural viewpoint, this term is analogous to the diamagnetic current term in the electrical conductivity~\cite{PhysRevB.99.045121,PhysRevB.108.085403}.

The third term $\kappa_{\perp}^{\nu;\rho}(i\omega_m)$ describes the ME response function defined entirely within the perpendicular complement ($p$-orbital) subspace given by
\begin{align}\begin{aligned}\label{eq:effKperp}
  & \kappa_{\perp}^{\nu;\rho}(i\omega_m) \\
  & \,\, = -\frac{\kB T}{V}\sum_{\epsilon_n} \Tr_{\perp} \left[ \hat{M}^\nu_{\perp} \hat{g}_{\perp}(i\epsilon_n + i\omega_m) \hat{J}^\rho_{\perp} \hat{g}_{\perp}(i\epsilon_n) \right],
\end{aligned}\end{align}
where $\Tr_{\perp}$ denotes the partial trace in $p$-orbital subspace $\mathbb{H}_{\perp}$.
The corresponding diagrams are shown in Fig.~\ref{fig:effsusdiag}(d).
This contribution vanishes when the chemical potential lies outside the $p$-orbital energy window: $\kappa_{\perp}^{\nu;\rho}(i\omega_m) = 0$.
Therefore, when focusing on low-energy physics near the $s$-orbital sector, the ME response function reduces to
\begin{align}
  \kappa^{\nu;\rho}(i\omega_m)
  =\chi^{\Pi^\nu;\Gamma^\rho}(i\omega_m) - \langle\Theta^{\nu;\rho}\rangle.
\end{align}

This result demonstrates that the finite ME response originates not from the effective Hamiltonian, but from vertex corrections and hybridization-induced terms that encode the influence of high-energy $p$-orbital processes.
Although the effective Hamiltonian correctly describes the low-energy sector, the ME response emerges only when the coupling to high-energy degrees of freedom is incorporated through these corrections, highlighting the crucial role of virtual $s$--$p$ hybridized processes.

\section{Renormalized Kubo Formula for the Effective Model} \label{sec:AC}

\subsection{Exact formulation of the renormalized Kubo formula} \label{sec:ACtrue}

To analyze the dynamical ME response in the real-frequency regime, we perform the analytic continuation of the Matsubara correlation functions via $i\omega_m \rightarrow \omega + i\gamma$. 
As a prerequisite, following the standard procedure used in the derivation of the Kubo formula (in the Lehmann representation)~\cite{PhysRev.118.1417,Altland_Simons_2010,PhysRevB.110.165111}, we first evaluate the sum over the fermionic Matsubara frequencies using the identity $\kB T\sum_{\epsilon_n}X(i\epsilon_n)=-\hbar\oint_{C}({\dd w}/{2\pi i}) f(\hbar w) X(w)$. 
Here, the contour $C$ encloses only the poles of the Fermi distribution function $f(\hbar w)$ while avoiding those of $X(w)$. 

After performing the analytic continuation, the ME response function is expressed as
\begin{widetext}
\begin{align}\label{eq:effMEKel}
  \kappa^{\nu;\rho}(\omega+i\gamma) 
  \cong -\frac{\hbar}{V} \int {\dd\Omega} f(\hbar\Omega)  \Tr_{\parallel}\left[ 
    \hat{\Pi}_{\mathrm{R}}^{\nu}\,\hat{\mathcal{G}}_{\mathrm{R}}(\Omega+\omega)\,\hat{\Gamma}_{\mathrm{R}}^{\rho}\,\hat{\mathcal{A}}(\Omega) + 
    \hat{\Pi}_{\mathrm{A}}^{\nu}\,\hat{\mathcal{A}}(\Omega)\,\hat{\Gamma}_{\mathrm{A}}^{\rho}\,\hat{\mathcal{G}}_{\mathrm{A}}(\Omega-\omega) + 
    \hat{\Theta}_{<}^{\nu;\rho}\,\hat{\mathcal{A}}(\Omega) 
  \right],
\end{align}
\end{widetext}
where
$\hat{\mathcal{G}}_{\mathrm{R}}(\Omega+\omega) = \hat{\mathcal{G}}_{\eff}(\Omega+\omega+i\gamma)$,
$\hat{\mathcal{G}}_{\mathrm{A}}(\Omega+\omega) = \hat{\mathcal{G}}_{\eff}(\Omega-\omega-i\gamma)$,
$\hat{\Pi}_{\mathrm{R}}^{\nu} = \hat{\Pi}^{\nu}(\Omega,\Omega+\omega+i\gamma)$,
$\hat{\Pi}_{\mathrm{A}}^{\nu} = \hat{\Pi}^{\nu}(\Omega-\omega-i\gamma,\Omega)$,
$\hat{\Gamma}_{\mathrm{R}}^{\rho} = \hat{\Gamma}^{\rho}(\Omega+\omega+i\gamma,\Omega)$, 
$\hat{\Gamma}_{\mathrm{A}}^{\rho} = \hat{\Gamma}^{\rho}(\Omega,\Omega-\omega-i\gamma)$, and $\hat{\Theta}_{<}^{\nu;\rho}=\hat{\Theta}^{\nu;\rho}(\Omega;\omega+i\gamma)$.
The $\delta$ function-like contribution from the perpendicular subspace ($\propto\delta(\hbar\Omega-\hat{h}_{\perp})$) is ignored, since it does not contribute in the low-energy region of interest.
The spectral function is defined as
\begin{align}\begin{aligned}
  \hat{\mathcal{A}}(k,\Omega)
  &=\frac{-1}{2\pi i}\left(\hat{\mathcal{G}}_{\eff}(k,\Omega+i0)-\hat{\mathcal{G}}_{\eff}(k,\Omega-i0)\right)\\
  &=\delta(\hbar\Omega-\hat{h}_{\eff}(k,\Omega)),
\end{aligned}\end{align}
which encodes the distribution of low-energy quasiparticle states dressed by high-energy contributions.

To evaluate the frequency integral and the trace in Eq.~(\ref{eq:effMEKel}), one needs to solve the self-consistent equation for the effective Hamiltonians:
\begin{align}\label{eq:SCeq}
  \hat{h}_{\eff}\left(k,\mathcal{E}_{k\mathfrak{a}}/\hbar\right) \ket{\mathfrak{u}_{k\mathfrak{a}}}= \mathcal{E}_{k\mathfrak{a}} \ket{\mathfrak{u}_{k\mathfrak{a}}}, 
\end{align}
where $\mathcal{E}_{k\mathfrak{a}}$ and $\ket{\mathfrak{u}_{k\mathfrak{a}}}$ represent the eigenvalues and eigenvectors of $\hat{h}_{\eff}\left(k,\mathcal{E}_{k\mathfrak{a}}/\hbar\right)$, respectively.
This corresponds to directly solving the Dyson equation self-consistently.
Using the solutions to this equation, the spectral function is decomposed as follows:
\begin{align}
  \hat{\mathcal{A}}(k,\Omega)
  =\sum_{\mathfrak{a}} Z_{k\mathfrak{a}} \delta(\hbar\Omega-\mathcal{E}_{k\mathfrak{a}}) \ket{\mathfrak{u}_{k\mathfrak{a}}}\bra{\mathfrak{u}_{k\mathfrak{a}}},
\end{align}
where $Z_{k\mathfrak{a}}=\bra{\mathfrak{u}_{k\mathfrak{a}}}1+\hat{\eta}^{\dagger}\hat{g}_{\perp}^2(\mathcal{E}_{k\mathfrak{a}}/\hbar)\hat{\eta}\ket{\mathfrak{u}_{k\mathfrak{a}}}^{-1}$ is the renormalization factor arising from $s$--$p$ hybridization.
This factor reflects the modification of spectral weight due to virtual processes involving high-energy $p$-orbital states.

We analytically solve the self-consistent equation in Eq.~(\ref{eq:SCeq}) and derive an explicit expression for the response functions.
For the projected effective Hamiltonian in Eq.~(\ref{eq:effhamdef}), 
the solutions can be obtained from those of the full Hamiltonian,
\begin{align}
  \hat{h}(k)\ket{u_{k\alpha}} = E_{k\alpha}\ket{u_{k\alpha}},
\end{align}
as~\cite{10.1063/1.1724312,RevModPhys.36.1076,LEINAAS197819,10.1063/1.4904200}
\begin{align}
  \mathcal{E}_{k\mathfrak{a}} = E_{k\mathfrak{a}}(\ll\Delta_p), \quad \ket{\mathfrak{u}_{k\mathfrak{a}}}=\frac{1}{\sqrt{Z_{k\mathfrak{a}}}}\hat{P}_{\parallel}\ket{u_{k\mathfrak{a}}},
\end{align}
where the eigenvalues of the effective Hamiltonian $\mathcal{E}_{k\mathfrak{a}}\,(\mathfrak{a}=1,\cdots,|\mathbb{H}_{\parallel}|/{N_\mathrm{k}})$ should be selected from those of the full Hamiltonian $E_{k\alpha}\,(\alpha=1,\cdots,|\mathbb{H}|/{N_\mathrm{k}})$ by identifying the states that are adiabatically connected to $\hat{h}_{\parallel}$, 
where $|\mathbb{H}_{\parallel}|=4N_{k}$ and $|\mathbb{H}|=16N_{k}$ denote the Hilbert-space dimensions.
When the $s$--$p$ energy separation $\Delta_p$ is sufficiently large, this condition corresponds to selecting the low-energy states satisfying $E_{k\mathfrak{a}} \ll \Delta_p$.
Using these states, the spectral function is expressed as
\begin{align}
  \hat{\mathcal{A}}(k,\Omega)=\sum_{\mathfrak{a}}^{E_{k\mathfrak{a}}\ll\Delta_p} \delta(\hbar\Omega-E_{k\mathfrak{a}}) \, \hat{P}_{\parallel}\ket{u_{k\mathfrak{a}}}\bra{u_{k\mathfrak{a}}}\hat{P}_{\parallel}.
\end{align}
This expression is consistent with the Schur-complement decomposition in Eq.~(\ref{eq:schur}).

Finally, substituting this expression into Eq.~(\ref{eq:effMEKel}), the ME response function is obtained as
\begin{widetext}
\begin{align}\label{eq:EffExSol}
  \kappa^{\nu;\rho}(\omega+i\gamma) 
  = -\frac{1}{V}\sum_{k}\sum_{\mathfrak{a}}^{E_{k\mathfrak{a}}\ll\Delta_p} f(E_{k\mathfrak{a}}) 
  \bra{u_{k\mathfrak{a}}} \Big( 
    \hat{\Pi}_{\mathrm{R}}^{\nu}\,\hat{\mathcal{G}}_{\mathrm{R}}(E_{k\mathfrak{a}}/\hbar+\omega)\,\hat{\Gamma}_{\mathrm{R}}^{\rho} + 
    \hat{\Gamma}_{\mathrm{A}}^{\rho}\,\hat{\mathcal{G}}_{\mathrm{A}}(E_{k\mathfrak{a}}/\hbar-\omega)\,\hat{\Pi}_{\mathrm{A}}^{\nu} + 
    \hat{\Theta}_{<}^{\nu;\rho} 
  \Big)\ket{u_{k\mathfrak{a}}}.
\end{align}
\end{widetext}
This expression explicitly shows that the dynamical ME response is governed by low-energy quasiparticle states, while the effects of high-energy $p$ orbitals are encoded through vertex corrections and the self-consistent renormalization.

\begin{figure*}[t]
  \centering
  \includegraphics[width=1.0\linewidth]{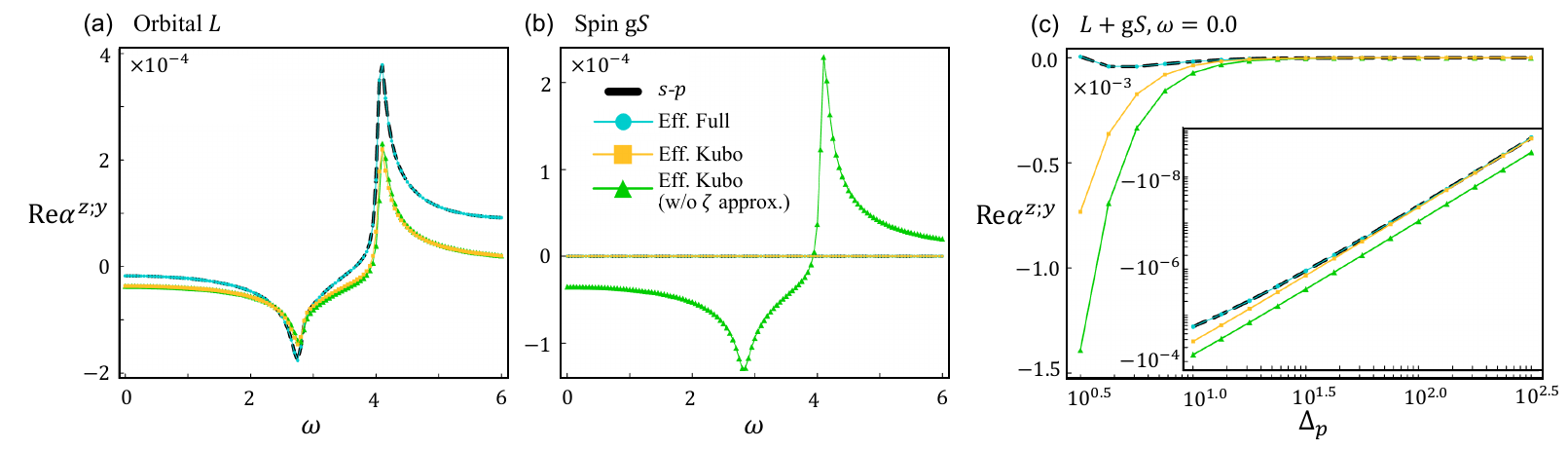}
  \caption{\label{fig:effres}
  Frequency $\omega$ dependence of the real part of the ME tensor component $\alpha^{z;y}(\omega)$ for (a) the atomic orbital magnetization and (b) the spin magnetization, obtained using different calculation methods. 
  In panel (a), the result obtained using Eq.~(\ref{eq:EffExSol}) (labeled ``Eff. Full''), which is calculated from the exact solutions with all perturbations included, perfectly reproduces the results of the $s$-$p$ model (labeled ``$s$-$p$''). 
  Furthermore, the results calculated using the corrected Kubo formula, Eq.~(\ref{eq:KuboMvRF}) (labeled ``Eff. Kubo''), adequately capture the qualitative behavior, including the peak structures. 
  The vicinity of the low-energy peak at $\omega=2.75$ is reproduced most accurately. 
  For comparison, the results obtained using the corrected Kubo formula without the $\zeta$ approximation (labeled ``Eff. Kubo (w/o $\zeta$ approx.)'') are also shown, yielding similar accuracy.
  In panel (b), only the result without the $\zeta$ approximation yields a finite value.
  (c) 
  $s$-$p$ energy separation $\Delta_p$ dependence of the ME tensor at $\omega=0$ for different calculation methods.
  In the large-$\Delta_p$ regime, incorporating the $\zeta$ approximation leads to convergence toward the results of the $s$--$p$ model, whereas a finite discrepancy remains without it.
  }
\end{figure*}

\subsection{Approximate evaluation with the renormalization factor} \label{sec:ACzeta}

The expression in Eq.~(\ref{eq:EffExSol}) requires solving a self-consistent equation or diagonalizing the full Hamiltonian, and further involves repeated evaluations of the Green's function within the summation, making it computationally demanding, especially for large matrix dimensions of the full Hamiltonian. 
To obtain a more tractable formulation, we introduce an approximate Green's function based on a low-energy expansion.

Expanding the effective Hamiltonian to linear order in $z$, we obtain:
\begin{align}
  \hat{h}_{\eff}(k,z)=\hat{h}_{\eff}(k,0)+(1-\hat{\zeta}^{-1})\hbar z + O\left( z^2 \hat{g}_{\perp}^2 \right),
\end{align}
where $\hat{\zeta}=\hat{\zeta}_k$ is a matrix-valued renormalization factor defined by
\begin{align}
  \hat{\zeta}^{-1}_k= 1+\hat{\eta}^{\dagger}\hat{g}^2_{\perp}(0)\hat{\eta}.
\end{align}
This approximation corresponds to retaining only the leading dynamical correction due to $s$--$p$ hybridization and can be regarded as a matrix-valued quasiparticle renormalization ($\zeta$ approximation).

Within this approximation, the Green's function is written as
\begin{align}\label{eq:GreeMvRF}\begin{aligned}
  \hat{\mathcal{G}}_{\eff}(k,z)  
  \cong \frac{1}{\hat{\zeta}^{-1}\hbar z-\hat{h}_{\eff}(k,0)}.
\end{aligned}\end{align}
This form can be diagonalized by solving the generalized eigenvalue problem
\begin{align}\label{eq:MvRFEvE}
  \hat{h}_{\eff}(k,0)\ket{\mathfrak{u}'_{k\mathfrak{a}}} 
  = \mathcal{E}'_{k\mathfrak{a}} \,\hat{\zeta}^{-1}\ket{\mathfrak{u}'_{k\mathfrak{a}}},
\end{align}
where $\mathcal{E}'_{k\mathfrak{a}}$ and $\ket{\mathfrak{u}'_{k\mathfrak{a}}}$ are the corresponding eigenvalues and eigenvectors, respectively. 
This leads to the spectral representation
\begin{align}
  \hat{\mathcal{G}}_{\eff}(k,z)
  = \sum_{\mathfrak{a}} 
  \frac{1}{\hbar z- \mathcal{E}'_{k\mathfrak{a}}} 
  \ket{\mathfrak{u}'_{k\mathfrak{a}}}\bra{\mathfrak{u}'_{k\mathfrak{a}}}\hat{\zeta}, 
\end{align}
where $\bra{\mathfrak{u}'_{k\mathfrak{a}}}$ is the left eigenvector: $\bra{\mathfrak{u}'_{k\mathfrak{a}}}\hat{\zeta}=\ket{\mathfrak{u}'_{k\mathfrak{a}}}^{\dagger}$.
The generalized eigenvalue problem has the same computational scaling as a standard eigenvalue problem~\cite{doi:10.1137/1.9781421407944}, making this approach numerically efficient.

The effective operators in $\zeta$ approximation are difined as follow:
\begin{subequations}\label{eq:effOszeta}\begin{align}
  \hat{\Pi}^{\prime\nu} 
  &= \hat{\Pi}^{\nu}(0,0),\\
  \hat{\Gamma}^{\prime\rho}(z_1,z_2)
  &=  \hat{\Gamma}^{\rho}(0,0) + \frac{z_1+z_2}{2}\Big.\frac{\partial\hat{\Gamma}^{\rho}(z,z)}{\partial z}\Big|_{z=0},\\
  \hat{\Theta}^{\prime\nu;\rho}
  &= \hat{\Theta}^{\nu;\rho}(0;0).
\end{align}\end{subequations}
These operators are constructed so as to satisfy the Ward--Takahashi identities associated with charge conservation and spin-$z$ conservation, 
ensuring consistency with the underlying symmetries of the system~\cite{PhysRev.78.182,takahashi1957generalized,TAYLOR1971436,JAHertz_1973,schrieffer1999theory,PhysRevD.101.074507,oliveira2023looking}.
In the present model, using $D\hat{P}_{\parallel}/Dk_y=0$, $[\hat{P}_{\parallel},\hat{s}^z]=0$, and $[\hat{h},\hat{s}^z]=0$, the operators reduce to
\begin{subequations}\begin{align}
  \hat{\Pi}^{\prime z}             &= \frac{\muB}{\hbar}\left(\mathrm{g}\hat{\zeta}^{-1}\hat{s}^{z}+\varDelta\hat{l}^{z}(0,0)\right),\\
  \hat{\Gamma}^{\prime y}(z_1,z_2) &= e\left(\frac{1}{\hbar}\frac{D \hat{h}_{\eff}(k,0)}{D k_y}-\frac{z_1+z_2}{2}\frac{D \hat{\zeta}^{-1}}{D k_y}\right),\\
  \hat{\Theta}^{\prime z;y}        &=\frac{e}{\hbar} \frac{D \hat{\Pi}^{\prime z}}{Dk_y}.
\end{align}\end{subequations}
Further details are provided in Appendix~\ref{sec:app:WTi}.

Using the representations of the Green's function in Eq.~(\ref{eq:GreeMvRF}) and the effective operators in Eq.~(\ref{eq:effOszeta}), 
the corrected Kubo formula incorporating both vertex corrections and the $\zeta$ approximation from $s$-$p$ mixing can be constructed:
\begin{align}\begin{aligned}\label{eq:KuboMvRF}
  \kappa^{\nu;\rho}(\omega+i\gamma)
  &\cong-\frac{1}{V}\sum_{k}\sum_{\mathfrak{a}\mathfrak{b}}\frac{f(\mathcal{E}'_{k\mathfrak{a}})-f(\mathcal{E}'_{k\mathfrak{b}})}{\mathcal{E}'_{k\mathfrak{a}}-\mathcal{E}'_{k\mathfrak{b}}+\hbar\omega+i\hbar\gamma}
  \Pi^{\nu}_{k;\mathfrak{a}\mathfrak{b}}\Gamma^{\rho}_{k;\mathfrak{b}\mathfrak{a}}\\
  &-\frac{1}{V}\sum_{k}\sum_{\mathfrak{a}} f(\mathcal{E}'_{k\mathfrak{a}})
  \left( {\Theta}^{\nu;\rho}_{k;\mathfrak{a}\mathfrak{a}} + \frac{1}{2\hbar} \{\Pi^{\nu},\partial\Gamma^{\rho}\}_{k;\mathfrak{a}\mathfrak{a}} \right),
\end{aligned}\end{align}
where
\begin{subequations}\label{eq:zetaope}\begin{align}
  \Pi^{\nu}_{k;\mathfrak{a}\mathfrak{b}}     
  &= \bra{\mathfrak{u}'_{k\mathfrak{a}}} \hat{\zeta}\,\hat{\Pi}^{\prime\nu}     \ket{\mathfrak{u}'_{k\mathfrak{b}}},\\
  \Gamma^{\rho}_{k;\mathfrak{b}\mathfrak{a}} 
  &= \bra{\mathfrak{u}'_{k\mathfrak{b}}} \hat{\zeta}\,\hat{\Gamma}^{\prime\rho}(\mathcal{E}'_{k\mathfrak{b}}/\hbar,\mathcal{E}'_{k\mathfrak{a}}/\hbar) \ket{\mathfrak{u}'_{k\mathfrak{a}}},
  \\
  {\Theta}^{\nu;\rho}_{k;\mathfrak{a}\mathfrak{a}}
  &= \bra{\mathfrak{u}'_{k\mathfrak{a}}} \hat{\zeta}\,\hat{\Theta}^{\prime\nu;\rho}\ket{\mathfrak{u}'_{k\mathfrak{a}}},
\end{align}\end{subequations}
and
\begin{align}\tag{\ref*{eq:zetaope}d}\begin{aligned}
  &\{\Pi^{\nu},\partial\Gamma^{\rho}\}_{k;\mathfrak{a}\mathfrak{a}}\\
  &\,\,= 2\Re\sum_{\mathfrak{b}}  \bra{\mathfrak{u}'_{k\mathfrak{a}}} \hat{\zeta}\,\hat{\Pi}^{\prime\nu} \ket{\mathfrak{u}'_{k\mathfrak{b}}}\bra{\mathfrak{u}'_{k\mathfrak{b}}} \hat{\zeta}\,\Big.\frac{\partial\hat{\Gamma}^{\rho}(z,z)}{\partial z}\Big|_{z=0} \ket{\mathfrak{u}'_{k\mathfrak{a}}}.
\end{aligned}\end{align}
This expression has the same computational structure as a conventional Kubo formula (the Lehmann representation), but incorporates the essential effects of $s$--$p$ hybridization through renormalized states and vertex corrections.
Thus, the dynamical ME response can be evaluated efficiently while retaining the key physics beyond the bare effective Hamiltonian.

It is noted that in our model, $D\hat{\zeta}^{-1}/Dk_y=0$. 
We also neglect the $\hat{\Theta}^{\prime z;y}$ term.
Consequently, the second term in Eq.~(\ref{eq:KuboMvRF}) can be omitted, and the frequency variable dependence of $\hat{\Gamma}^{\prime y}$ are not considered.

\section{Discussion} \label{sec:disc}

\subsection{Numerical results based on the effective model} \label{subsec:conpairapp}

\begin{figure*}[t]
  \centering
  \includegraphics[width=1.0\linewidth]{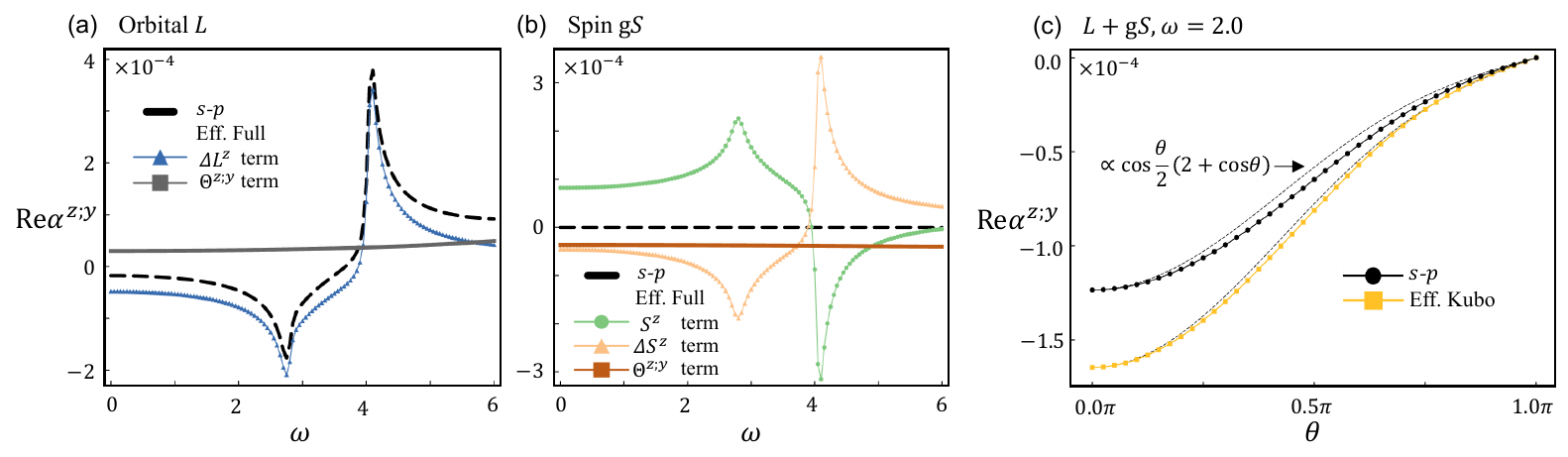}
  \caption{\label{fig:contri}
  [(a) and (b)] 
  Frequency $\omega$ dependence of the real part of the ME tensor component $\alpha^{z;y}(\omega)$, decomposed into contributions from (a) the atomic orbital magnetization and (b) the spin magnetization.
  The overall $\omega$ dependence in (a) is dominated by the orbital magnetization term (labeled ``$\varDelta L^z$ term''). 
  In contrast, the magnetoelectric operator term (labeled ``$\Theta^{z;y}$ term'') exhibits only a weak frequency dependence and remains nearly constant.
  For the spin magnetization in panel (b), the individual contributions cancel each other, resulting in a vanishing total response. 
  (c)
  Zigzag apex angle $\theta$ dependence of the ME tensor at $\omega=2.0$. 
  The dashed line represents the function $\cos(\theta/2)(2+\cos\theta)$ scaled to match the data at $\theta=0$.
  }
\end{figure*}

We evaluate the frequency $\omega$ dependence of the ME tensor component $\alpha^{z;y}$ as Eq.~(\ref{eq:spsusG}) in the effective model. 
Figure~\ref{fig:effres}(a) shows the real part of the atomic orbital ($L$) magnetization contributions obtained by different calculation methods.
As shown in Fig.~\ref{fig:effres}(a), the result based on Eq.~(\ref{eq:EffExSol}) (``Eff. Full''), which incorporates the exact solutions of the eigenvalue equation including all perturbative effects, quantitatively reproduces the result of the original $s$--$p$ model.
This demonstrates that the ME response can be faithfully recovered within the effective model, provided that the effects of the eliminated $p$-orbital degrees of freedom are properly incorporated through vertex corrections and renormalization effects.

Figure~\ref{fig:effres}(a) also shows the results calculated using the corrected Kubo formula, Eq.~(\ref{eq:KuboMvRF}) (``Eff. Kubo'').
This approximation captures the overall frequency dependence, including the peak structures, in good agreement with the exact result.
In particular, the low-energy peak around $\omega=2.75$ is reproduced with high accuracy. 
This can be understood from the fact that this peak is dominated by transitions near the Fermi level, where the $\zeta$ approximation based on a low-energy expansion is most reliable. 
For comparison, we also show the results obtained without solving the generalized eigenvalue problem in Eq.~(\ref{eq:MvRFEvE}), i.e., without incorporating the $\zeta$ approximation.
A comparable level of accuracy is obtained, indicating that the qualitative features of the ME response are robust against this approximation.

Figure~\ref{fig:effres}(b) shows the spin ($\mathrm{g}S$) magnetization contributions for different calculation methods.
Although the effective Hamiltonian preserves $S^z$ and hence the spin contribution is expected to vanish, a finite value appears when the $\zeta$ approximation is not included.
This spurious contribution originates from the violation of the Ward--Takahashi identity between the effective operators, reflecting the breakdown of the underlying conservation laws at the level of the approximate theory.
In contrast, incorporating the $\zeta$ approximation restores this identity and correctly suppresses the spin contribution.

Figure~\ref{fig:effres}(c) shows the $s$-$p$ energy separation $\Delta_p$ dependence of the ME tensor at $\omega=0$ for different calculation methods.
With the $\zeta$ approximation, the results converge to those of the $s$-$p$ model in the limit of large $s$-$p$ energy separation, as expected when the low-energy $s$-orbital subspace becomes well separated from the high-energy $p$-orbital states. 
In contrast, without the $\zeta$ approximation, a finite spurious contribution to the spin magnetization persists, and the results fail to converge even in this limit.
This clearly demonstrates that the $\zeta$ approximation is essential for correctly incorporating the effects of virtual $s$--$p$ processes and ensuring consistency with the full model.

The results of numerical analytic continuation using the barycentric rational function approximation~\cite{PhysRevB.111.125139} are presented in Appendix~\ref{sec:app:ACFlow}.

\subsection{Contribution decomposition} \label{subsec:contri}

We analyze the $\omega$ dependence of the ME tensor by decomposing the contributions from each term in the effective magnetization operator $\hat{\Pi}^{z}$ and the effective magnetoelectric operator $\hat{\Theta}^{z;y}$, based on the exact solution [Eq.~(\ref{eq:EffExSol})].

Figure~\ref{fig:contri}(a) shows the decomposition of the atomic orbital magnetization contribution.
The overall $\omega$ dependence is governed by the orbital magnetization term ($\varDelta L^z$;$\Gamma^y$) in the corrected magnetization--current correlation function. 
In contrast, the magnetoelectric
operator term ($\Theta^{z;y}$) exhibits only a weak dependence on $\omega$ and remains nearly constant over the relevant frequency range.
This behavior can be understood from its origin: the frequency dependence of $\Theta^{z;y}$ arises through the propagator $\hat{g}_{\perp}(z)$, 
whose contribution is suppressed by the large $s$--$p$ energy separation $\Delta_p$ in the energy window dominated by the $s$-orbital states.
Thus, the dynamical ME response is primarily governed by interband processes encoded in the vertex-corrected orbital magnetization term.

Figure~\ref{fig:contri}(b) shows the decomposition of the spin magnetization contribution.
Although individual terms yield finite contributions, they cancel each other, resulting in a vanishing total response.
This cancellation reflects the conservation of the $z$-component of spin and the associated Ward--Takahashi identity, which enforces the absence of a spin contribution to the ME response.
The additional contributions arising from the effective operators ($\varDelta S^z$;$\Gamma^y$) exhibit a finite structure similar to that observed in Fig.~\ref{fig:effres}(b).
However, these contributions are exactly canceled by other terms in the full expression.
In contrast, within the Kubo formula without the $\zeta$ approximation, the compensating terms are absent, leading to an incomplete cancellation and hence a spurious finite spin contribution.

\subsection{Essential model parameters} \label{subsec:selrule}

In this section, we identify the essential model parameters governing the ME tensor $\alpha^{z;y}$ in the effective model.
To this end, we analyze the following trace quantities obtained from the expansion of the ME tensor~\cite{doi:10.7566/JPSJ.91.014701,PhysRevB.105.155157,PhysRevB.107.155109}:
\begin{align}
  \label{eq:EPdef1}|\mathbb{H}|^{-1}\,          \Tr\,          &\left[\hat{M}^{z}\,(\hat{h})^{n_1}  \hat{J}^{y}\,(\hat{h})^{n_2}\right],\\
  \label{eq:EPdef2}|\mathbb{H}_{\parallel}|^{-1}\Tr_{\parallel}&\left[\hat{\zeta}\hat{\Pi}^{z}\,(\hat{\zeta}\hat{h}_{\eff})^{n_1}\,\hat{\zeta}\hat{\Gamma}^{y}\,(\hat{\zeta}\hat{h}_{\eff})^{n_2}\right],\\
  \label{eq:EPdef3}|\mathbb{H}_{\parallel}|^{-1}\Tr_{\parallel}&\left[\hat{\Pi}^{z}\,(\hat{h}_{\eff})^{n_1}\,\hat{\Gamma}^{y}\,(\hat{h}_{\eff})^{n_2}\right],
\end{align}
which correspond to the $s$--$p$ model, the corrected Kubo formula with the $\zeta$ approximation, and the same formula without the $\zeta$ approximation, respectively.
Here, $n_1$ and $n_2$ are integers, 
$|\mathbb{H}|=16N_{k}$ and $|\mathbb{H}_{\parallel}|=4N_{k}$ denote the Hilbert-space dimensions, and the frequency variable of the effective Hamiltonian and effective operators is set to $z=0$.
For simplicity, we set $e=\hbar=\muB=d=1$, $\mathrm{g}=2$, and replace the $k$-summation by an integral: $(1/N_k)\sum_{k}\rightarrow\int^{\pi
}_{-\pi}{\dd (ka)/2\pi}$.

\paragraph*{Orbital magnetization contribution.}
First, we consider the orbital magnetization term.
In the effective model, the correction to the orbital magnetization operator is expressed as follows:
\begin{align}\begin{aligned}
  \varDelta \hat{l}^z 
  &    = l_1 (1+\cos\theta\cos ka)(\hat{\tau}_3\hat{\sigma}_0)+ l_2 \sin\theta\sin ka (\hat{\tau}_3\hat{\sigma}_0)\\
  &\,\,+ l_3 (1+\cos\theta\cos ka)(\hat{\tau}_0\hat{\sigma}_z)+ l_4 \sin\theta\sin ka (\hat{\tau}_0\hat{\sigma}_z),
\end{aligned}\end{align}
where, under the condition $\Delta_p\rightarrow\infty$, $l_1$--$l_4$ denote
\begin{align}\begin{aligned}
  l_1=l_4\cong \frac{4t_{sp\sigma}^2\phi  }{\Delta_p^3},\,
  l_2    \cong \frac{2t_{sp\sigma}^2      }{\Delta_p^2},\,
  l_3    \cong \frac{12t_{sp\sigma}^2\phi^2}{\Delta_p^4}.
\end{aligned}\end{align}
The lowest-order contribution to $\alpha^{z;y}$ arises from $(n_1,n_2)=(1,0)$, yielding
\begin{align}
  &\begin{aligned}
    &|\mathbb{H}_{\parallel}|^{-1} \Tr_{\parallel}\left[\hat{\zeta}\varDelta\hat{l}^z\,\hat{\zeta}\hat{h}_{\eff}\,\hat{\zeta}\hat{\Gamma}^{y}\right]\\
    &\qquad\qquad= i\,t_{ss\sigma}^2 l_1  \cos\frac{\theta}{2} (2+\cos\theta)+O(\Delta_p^{-5}),
  \end{aligned}\\ \notag \\
  &\begin{aligned}
    &|\mathbb{H}_{\parallel}|^{-1} \Tr_{\parallel}\left[\varDelta\hat{l}^z\,\hat{h}_{\eff}\,\hat{\Gamma}^{y}\right]\\
    &\qquad\qquad=     i\,t_{ss\sigma}^2 l_1 \cos\frac{\theta}{2} (2+\cos\theta)\\
    &\qquad\qquad\cong 4i\frac{t_{ss\sigma}^2t_{sp\sigma}^2\phi}{\Delta_p^3} \cos\frac{\theta}{2} (2+\cos\theta).
  \end{aligned}
\end{align}
These results indicate that, in the orbital magnetization term, the effect of the renormalization factor is practically negligible ($\hat{\zeta}\cong1$), which is consistent with the observation that the results exhibit nearly identical behavior in Fig.~\ref{fig:effres}(a).
The dominant contribution originates from the $l_1$ term, which describes the coupling between the orbital magnetization and the AFM molecular field. 
This term can be interpreted as a virtual $s$--$p$ hybridization process:
\begin{align}
  l_1 (1+\cos\theta\cos ka)(\hat{\tau}_3\hat{\sigma}_0)\cong - \hat{\eta}^{\dagger} \frac{1}{\Delta_p}\,\hat{l}^z\,\frac{\phi}{\Delta_p^2}(\hat{l}^z\hat{\tau}_3)\,\hat{\eta}.
\end{align}
This indicates that the ME response arises from higher-order hybridization processes involving both orbital angular momentum and the molecular field. 
In combination with $s$-orbital hopping and the current operator, this process generates a finite trace contribution.

Meanwhile, in the full $s$--$p$ model, the essential contribution appears at higher order $(n_1,n_2)=(3,2)$:
\begin{align}\begin{aligned}
  &|\mathbb{H}|^{-1}\Tr\left[\hat{M}^{z}\,(\hat{h})^{3}\hat{J}^{y}\,(\hat{h})^{2}\right]\\
  &\quad=    - 2i\,t_{ss\sigma}^2 t_{sp\sigma}^2 \phi \Delta_p \cos\frac{\theta}{2} (2+\cos\theta)\\
  &\quad\qquad -i\,t_{sp\sigma}^2 \phi \Delta_p \left[\phi^2 + t_{sp\sigma}^2 (2+\cos2\theta)\right]  \cos\frac{\theta}{2} .
\end{aligned}\end{align}
In this way, the effective model partially reproduces the results of the $s$-$p$ model except for the $\Delta_p$ dependence. 
The second term represents transition processes that do not involve the $s$-orbital, and is therefore not captured within the effective model. 
These contributions become relevant only when the chemical potential lies in the $p$-orbital energy region.

The resulting angular dependence $\cos(\theta/2)(2+\cos\theta)$ agrees well with the numerical results shown in Fig.~\ref{fig:contri}(c).
This agreement confirms that the leading-order processes identified in the above analysis capture the essential physics of the ME response.

\paragraph*{Spin magnetization contribution.}
Next, we consider the spin magnetization term. 
When spin-$z$ conservation is preserved, all contributions cancel exactly:
\begin{align}
  |\mathbb{H}|^{-1}\Tr\left[\hat{s}^{z}\,(\hat{h})^{n_1}  \hat{J}^{y}\,(\hat{h})^{n_2}\right]
  &=0,\\
  |\mathbb{H}_{\parallel}|^{-1}\Tr_{\parallel}\left[\hat{\zeta}(\hat{\zeta}^{-1}\hat{s}^{z})\,(\hat{\zeta}\hat{h}_{\eff})^{n_1}\,\hat{\zeta}\hat{\Gamma}^{y}\,(\hat{\zeta}\hat{h}_{\eff})^{n_2}\right]
  &=0.
\end{align}
This cancellation reflects the conservation of $S^z$ and the associated Ward--Takahashi identity.

However, without the $\zeta$ approximation, the cancellation is incomplete.
At the lowest order $(n_1,n_2)=(1,0)$, a finite contribution remains:
\begin{align}\begin{aligned}
  |\mathbb{H}_{\parallel}|^{-1} \Tr_{\parallel}\left[(\hat{\zeta}^{-1}\hat{s}^{z})\,\hat{h}_{\eff}\,\hat{\Gamma}^{y}\right]
  &=  -\frac{3}{2}i\,t_{ss\sigma}^2 \frac{\partial \phi'}{\partial z} \cos\frac{\theta}{2}\\
  &\cong 6i \frac{t_{ss\sigma}^2t_{sp\sigma}^2\phi}{\Delta_p^3}\cos\frac{\theta}{2}.
\end{aligned}\end{align}
This spurious contribution originates from the frequency dependence of the effective spin operator through $\hat{\zeta}^{-1}$, specifically via the term $\partial \phi'/\partial z$.
Without the $\zeta$ approximation, the corresponding counterterms required by the Ward--Takahashi identity are missing, and the cancellation is violated. 
Since this contribution is of the same order $O(\Delta_p^{-3})$ as the orbital term, it leads to a significant error in the ME response.

These results demonstrate that the $\zeta$ approximation is essential not only for quantitative accuracy but also for preserving the symmetry constraints of the theory.
It plays a crucial role in enforcing conservation laws, such as spin-$z$ conservation, and eliminating spurious contributions that would otherwise lead to qualitatively incorrect ME responses.

\section{Summary} \label{sec:Summary}

In summary, we have theoretically investigated the ME effect in the AFM zigzag chain.
We first evaluated the frequency dependence of the ME response using the Kubo formula in the $s$--$p$ orbital model.
The numerical results reveal that only the orbital magnetization contribution remains finite, while the spin magnetization contribution vanishes, reflecting the conservation of the $z$-component of spin in the model.
This demonstrates that the ME effect in the present system is governed by orbital degrees of freedom rather than spin.

We then examined whether this ME effect can be captured within a low-energy effective Hamiltonian restricted to the $s$-orbital subspace.
By constructing the effective Hamiltonian via projection, we found that it is essentially equivalent to that used in previous studies, and that a naive application of the Kubo formula fails to produce a finite ME response.
This highlights a fundamental limitation of conventional low-energy descriptions.
To address this issue, we reformulated the ME tensor using the Schur complement and derived an effective description in which vertex corrections and hybridization-induced terms are explicitly incorporated.
Within this framework, the effects of orbital magnetization appear through vertex-corrected operators, which encode virtual processes involving the eliminated $p$-orbital degrees of freedom. 
Furthermore, by introducing the renormalization factor ($\zeta$) approximation, we derive a corrected Kubo formula that consistently incorporates both hybridization effects and the underlying conservation laws.

Using this formalism, we performed numerical calculations within the effective model.
We showed that, when all corrections are properly included, the effective model quantitatively reproduces the results of the original $s$--$p$ model.
Moreover, we demonstrated that approximations such as vertex corrections and the $\zeta$ approximation are crucial for preserving spin conservation and accurately describing the orbital contribution.
A decomposition of the ME response further clarified that the dominant contribution arises from the orbital magnetization term, while the spin contribution vanishes due to symmetry constraints.
By analyzing the essential model parameters, we identified the key processes responsible for the emergence of the ME effect and clarified the role of selection rules and hybridization.

Possible candidate materials for realizing the ME effect discussed in this study include $f$-electron systems with strong orbital character and parity-mixed hybridization.
These can be broadly classified into two categories. 
The first category consists of Ce-based intermetallic compounds characterized by zigzag-like lattice structures, 
such as 
CeRu$_2$Al$_{10}$~\cite{Tursina:wm6046,doi:10.1143/JPSJ.80.073701,PhysRevB.85.205208,Adroja_2013}, 
CeOs$_2$Al$_{10}$~\cite{doi:10.1143/JPSJ.80.073701,PhysRevB.85.205208,Adroja_2013}, 
Ce$_3$TiBi$_5$~\cite{MOTOYAMA2018142,doi:10.7566/JPSCP.30.011189,doi:10.7566/JPSJ.89.033703}, and 
Ce$_3$ZrBi$_5$~\cite{doi:10.7566/JPSCP.30.011180}. 
The second category includes related low-dimensional $f$-electron systems, such as 
$\alpha$-YbAl$_{1-x}$Mn$_x$B$_4$~\cite{PhysRevResearch.3.023140} 
and NdRu$_2$Al$_{10}$~\cite{13pd-tlzp}, which exhibit strong parity-mixed $c$--$f$ hybridization and anisotropic magnetic interactions. 
In these systems, the coexistence of parity-mixed hybridization and AFM order serves as the key ingredients for activating the orbital contribution emphasized in our proposed mechanism.
Although finite interchain hopping is generally expected in real materials and may quantitatively modify the ME response, we expect the essential mechanism discussed here to remain operative as long as the locally noncentrosymmetric structure and the symmetry allowing the ME response are preserved.

Finally, the methodology developed in this study, namely, the construction of effective Hamiltonians and the systematic incorporation of vertex corrections in response functions, provides a revisited framework for low-energy theories, in which hybridization effects beyond conventional descriptions are consistently included. 
This approach can be broadly applied to systems partitioned into low-energy and high-energy subspaces. 
This framework may provide a general approach for incorporating orbital hybridization effects beyond simple low-energy models. 
It can be extended to first-principles calculations and Wannier-based model construction~\cite{PhysRevB.65.035109, mostofi2008wannier90, Ozaki_CW_PRB_2024, oiwa2025symmetry, kirikoshi2026nonlocal}, as well as to spatially inhomogeneous systems such as open or interface systems.

\begin{acknowledgments}
We thank Junya Otsuki, Akimitsu Kirikoshi, Rikuto Oiwa, Yuuki Ogawa, Takahiro Morimoto, and Sota Kitamura for helpful discussions and comments.
This research was supported by 
JST SPRING (JPMJSP2119), 
JST BOOST (JPMJBS2426), 
JSPS KAKENHI (JP22H00101, JP23H04869), and by 
JST CREST (JPMJCR23O4) and 
JST FOREST (JPMJFR2366).
\end{acknowledgments}

\section*{Data Availability}

The data that support the findings of this article are openly available in Zenodo at \cite{dataset_zenodo}.

\appendix

\begin{widetext}
\section{Full expressions for the effective operaters} \label{sec:app:Operators}

The full expressions for the effective operators used in Eq.~(\ref{eq:effsus}), as illustrated in Figs.~\ref{fig:effsusdiag}(c) and (e), are given by
\begin{subequations}\begin{align}
  \hat{\Pi}^{\nu}(z_1,z_2)     
  &= \hat{M}^{\nu}_{\parallel}
  \,+\,  \hat{\eta}^{\dag} \hat{g}_{\perp}(z_1) \hat{M}^{\nu}_{\perp} \hat{g}_{\perp}(z_2) \hat{\eta}
  \,+\, \left( 
    \hat{m}^{\nu\dag} \hat{g}_{\perp}(z_2) \hat{\eta}
    \,+\, \hat{\eta}^{\dag} \hat{g}_{\perp}(z_1) \hat{m}^{\nu}
  \right),\\
  \hat{\Gamma}^{\rho}(z_1,z_2) 
  &= \hat{J}^{\rho}_{\parallel}
  \,+\,  \hat{\eta}^{\dag} \hat{g}_{\perp}(z_1) \hat{J}^{\rho}_{\perp} \hat{g}_{\perp}(z_2) \hat{\eta}
  \,+\, \left( 
    \hat{j}^{\rho\dag} \hat{g}_{\perp}(z_2) \hat{\eta}
    \,+\, \hat{\eta}^{\dag} \hat{g}_{\perp}(z_1) \hat{j}^{\rho}
  \right),\\
  \hat{\Theta}^{\nu;\rho}(z;z')
    &=\begin{aligned}[t]
    &\left( 
      \hat{m}^{\nu\dag} \hat{g}_{\perp}(z+z') \hat{j}^{\rho} 
      + \hat{j}^{\rho\dag} \hat{g}_{\perp}(z-z') \hat{m}^{\nu}
    \right)
    + \left( 
      \hat{\eta}^{\dag} \hat{g}_{\perp}(z) \hat{M}^{\nu}_{\perp} \hat{g}_{\perp}(z+z') \hat{J}^{\rho}_{\perp}  \hat{g}_{\perp}(z) \hat{\eta}
      + \hat{\eta}^{\dag} \hat{g}_{\perp}(z) \hat{J}^{\rho}_{\perp} \hat{g}_{\perp}(z-z') \hat{M}^{\nu}_{\perp} \hat{g}_{\perp}(z) \hat{\eta}
    \right)\\
    &\,+\left( 
      \hat{\eta}^{\dag} \hat{g}_{\perp}(z) \hat{M}^{\nu}_{\perp} \hat{g}_{\perp}(z+z') \hat{j}^{\rho}  
      + \hat{j}^{\rho\dag}  \hat{g}_{\perp}(z-z') \hat{M}^{\nu}_{\perp} \hat{g}_{\perp}(z) \hat{\eta}
    \right)
    + \left( 
      \hat{m}^{\nu\dag} \hat{g}_{\perp}(z+z') \hat{J}^{\rho}_{\perp} \hat{g}_{\perp}(z) \hat{\eta}
      + \hat{\eta}^{\dag} \hat{g}_{\perp}(z) \hat{J}^{\rho}_{\perp} \hat{g}_{\perp}(z-z') \hat{m}^{\nu}
    \right),
  \end{aligned}
\end{align}\end{subequations}
where the operators are decomposed into the respective subspaces as follows:
\begin{align}
  \hat{h} = \begin{pmatrix}
    {\hat{h}_{\parallel}} & {\hat{\eta}^{\dagger}} \\ {\hat{\eta}} & {\hat{h}_{\perp}}
  \end{pmatrix},\qquad
  \hat{M}^{\nu} = \begin{pmatrix}
    {\hat{M}^{\nu}_{\parallel}} & {\hat{m}^{\nu\dag}} \\ {\hat{m}^{\nu}} & {\hat{M}^{\nu}_{\perp}}
  \end{pmatrix} ,\qquad
  \hat{J}^{\rho} = \begin{pmatrix}
    {\hat{J}^{\rho}_{\parallel}} & {\hat{j}^{\rho\dag}} \\ {\hat{j}^{\rho}} & {\hat{J}^{\rho}_{\perp}}
  \end{pmatrix}.
\end{align}
For our model in Eq.~(\ref{eq:spHam}), these matrix elements satisfy $\hat{m}^{\nu}=0$ and $\hat{J}^{\rho}_{\perp}=0$.

\end{widetext}
\section{Ward--Takahashi identity for the effective operators} \label{sec:app:WTi}

In this Appendix, we demonstrate that the effective operators introduced in the main text satisfy the Ward--Takahashi identity~\cite{PhysRev.78.182,takahashi1957generalized,TAYLOR1971436,JAHertz_1973,schrieffer1999theory,PhysRevD.101.074507,oliveira2023looking}.
This identity ensures that the effective theory respects the underlying conservation laws, such as spin and charge conservation, even after integrating out the high-energy degrees of freedom.
For convenience, we introduce the following unitary transformation for each operator: $\breve{\sharp}(\vb*{k}_1,\vb*{k}_2)=e^{-i\vb*{k}_1\cdot\vb*{\hat{d}}}\,\hat{\sharp}\,e^{i\vb*{k}_2\cdot\vb*{\hat{d}}}$. 
Under this transformation, the electric current operator, modified from Eq.~(\ref{eq:Jmatrix}),
is expressed as $\vb*{\breve{J}} = ({e}/{\hbar})({\partial \breve{h}(\vb*{k})}/{\partial \vb*{k}})$.

\paragraph*{Spin magnetization operator.}
First, we consider the Ward--Takahashi identity associated with the $z$-spin $U(1)$ symmetry.
The identity is given by
\begin{align}\begin{aligned}
  (\hbar z_1-\hbar z_2)\breve{\Gamma}_{s^z}^0&(\vb*{k}_1z_1,\vb*{k}_2z_2)-(\hbar\vb*{k}_1-\hbar\vb*{k}_2)\cdot\breve{\vb*{\Gamma}}_{s^z} (\vb*{k}_1z_1,\vb*{k}_2z_2)\\
  &=\breve{\mathcal{G}}^{-1}_{\eff}(\vb*{k}_1,z_1)\hat{s}^z-\hat{s}^z\breve{\mathcal{G}}_{\eff}^{-1}(\vb*{k}_2,z_2),
\end{aligned}\end{align}
where $\breve{\Gamma}_{s^z}^0$ and $\breve{\vb*{\Gamma}}_{s^z}$ are the effective local spin density and current operators, respectively.
The effective magnetization operator $\hat{\Pi}^{z}$ satisfies the following relation:
\begin{align}
  \hat{\Pi}^{z}(z_1,z_2)=\frac{\muB}{\hbar}\left( e^{i\vb*{k}\cdot\vb*{\hat{d}}} \mathrm{g}\breve{\Gamma}_{s^z}^0(\vb*{k}z_1,\vb*{k}z_2) e^{-i\vb*{k}\cdot\vb*{\hat{d}}} +\varDelta l^z (z_1,z_2) \right).
\end{align}
Since neither the spin current nor the spin torque is generated ($\breve{\vb*{\Gamma}}_{s^z}=\vb*{0}$), it is sufficient to consider the limit $\vb*{k}_1-\vb*{k}_2\rightarrow\vb*{0}$. 
Using $[\hat{P}_{\parallel},\hat{s}^z]=0$ and $[\hat{h},\hat{s}^z]=0$, the Ward--Takahashi identity is given by
\begin{align}\label{eq:WTIspin}
  (\hbar z_1-\hbar z_2)\,\breve{\Gamma}_{s^z}^0(\vb*{k}z_1,\vb*{k}z_2)=\hat{s}^z\left(\breve{\mathcal{G}}^{-1}_{\eff}(\vb*{k},z_1)-\breve{\mathcal{G}}_{\eff}^{-1}(\vb*{k},z_2)\right).
\end{align}
Evaluating the right-hand side explicitly, we obtain
\begin{align}\begin{aligned}
  \hat{s}^z&\left(\breve{\mathcal{G}}^{-1}_{\eff}(\vb*{k},z_1)-\breve{\mathcal{G}}_{\eff}^{-1}(\vb*{k},z_2)\right)\\
  &=\hat{s}^z\left[(\hbar z_1-\hbar z_2)-\left(\breve{h}(\vb*{k},z_1)-\breve{h}(\vb*{k},z_2)\right)\right]\\
  &=(\hbar z_1-\hbar z_2)\hat{s}^z\left(1+\breve{\eta}^{\dagger}\breve{g}_{\perp}(z_1)\breve{g}_{\perp}(z_2)\breve{\eta}\right)\\
  &=(\hbar z_1-\hbar z_2)\left(\hat{s}^z+\varDelta \breve{s}^z(z_1,z_2)\right).
\end{aligned}\end{align}
This agrees with the result derived from the effective operator transformation according to Eq.~(\ref{eq:effMag}), 
demonstrating that the effective operator in Eq.~(\ref{eq:effMag}) satisfies the Ward--Takahashi identity.

When the $\zeta$ approximation is applied, the right-hand side of Eq.~(\ref{eq:WTIspin}) becomes
\begin{align}\begin{aligned}
  \hat{s}^z&\left(\breve{\mathcal{G}}^{-1}_{\eff}(\vb*{k},z_1)-\breve{\mathcal{G}}_{\eff}^{-1}(\vb*{k},z_2)\right)
  \cong (\hbar z_1-\hbar z_2) \, \breve{\zeta}^{-1}\hat{s}^z,
\end{aligned}\end{align}
which demonstrates that the effective operator defined in Eq.~(\ref{eq:effOszeta}) also satisfies the Ward--Takahashi identity.

\paragraph*{Current operator.}
Next, we consider the Ward--Takahashi identity associated with charge $U(1)$ symmetry.
The identity is given by
\begin{align}\begin{aligned}\label{eq:A6}
  (\hbar z_1-\hbar z_2)\,\breve{\Gamma}_{e}^0&(\vb*{k}_1z_1,\vb*{k}_2z_2)-(\hbar\vb*{k}_1-\hbar\vb*{k}_2)\cdot\breve{\vb*{\Gamma}}_{e} (\vb*{k}_1z_1,\vb*{k}_2z_2)\\
  & =e\left(\breve{\mathcal{G}}^{-1}_{\eff}(\vb*{k}_1,z_1)-\breve{\mathcal{G}}_{\eff}^{-1}(\vb*{k}_2,z_2)\right),
\end{aligned}\end{align}
where $\breve{\Gamma}_{e}^0$ and $\breve{\vb*{\Gamma}}_{e}$ are the effective local electric density and current operators, respectively.
The effective current operator $\hat{\Gamma}^{\rho}$ introduced above satisfies the following relation:
\begin{align}
  \hat{\Gamma}^{\rho}(z_1,z_2)=e^{i\vb*{k}\cdot\vb*{\hat{d}}} \breve{\Gamma}^{\rho}_{e}(\vb*{k}z_1,\vb*{k}z_2) e^{-i\vb*{k}\cdot\vb*{\hat{d}}},
\end{align}
using $D\hat{P}_{\parallel}/Dk_y=0$.
To obtain the electric current operator under the uniform expansion, 
by expanding up to the linear order in $\vb*{q}=\vb*{k}_1-\vb*{k}_2$ and comparing both sides, 
the expression in Eq.~(\ref{eq:A6}) is rewritten as follows:
\begin{align}\begin{aligned}
  (\hbar &z_1-\hbar z_2)\,\breve{\Gamma}_{e}^0(\vb*{k}z_1,\vb*{k}z_2)\\
  & -\hbar\vb*{q}\cdot\left(\breve{\vb*{\Gamma}}_{e} (\vb*{k}z_1,\vb*{k}z_2)-\left.(z_1-z_2)\frac{\partial\breve{\Gamma}_{e}^0(\vb*{k}'z_1,\vb*{k}z_2)}{\partial \vb*{k}'}\right|_{\vb*{k}'\rightarrow\vb*{k}}\right)\\
  & =e\left(\breve{\mathcal{G}}^{-1}_{\eff}(\vb*{k},z_1)-\breve{\mathcal{G}}_{\eff}^{-1}(\vb*{k},z_2)+\vb*{q}\cdot\frac{\partial \breve{\mathcal{G}}^{-1}_{\eff}(\vb*{k},z_1)}{\partial \vb*{k}} \right).
\end{aligned}\end{align}
By applying (anti-)symmetrization under the exchange of $\vb*{k}_1$($z_1$) and $\vb*{k}_{2}$($z_2$), the effective charge density operator can be obtained as follows:
\begin{align}
  \breve{\Gamma}_{e}^0(\vb*{k}_1z_1,\vb*{k}_2z_2) &= e\left(1+\breve{\eta}_{\vb*{k}_1}^{\dagger}\breve{g}_{\perp}(\vb*{k}_1,z_1)\breve{g}_{\perp}(\vb*{k}_2,z_2)\breve{\eta}_{\vb*{k}_2}\right).
\end{align}
Therefore, we obtain
\begin{subequations}\begin{align}
  &\breve{\Gamma}_{e}^0(\vb*{k}z_1,\vb*{k}z_2) 
  = e\left(1+\breve{\eta}^{\dagger}\breve{g}_{\perp}(z_1)\breve{g}_{\perp}(z_2)\breve{\eta}\right),\\
  &\left.\frac{1}{\hbar}\frac{\partial\breve{\Gamma}_{e}^0(\vb*{k}'z_1,\vb*{k}z_2)}{\partial \vb*{k}'}\right|_{\vb*{k}'\rightarrow\vb*{k}}\\
  &\qquad= \notag
    \breve{\vb*{j}}^{\dagger}\breve{g}_{\perp}(z_1)\breve{g}_{\perp}(z_2)\breve{\eta}+
    \breve{\eta}^{\dagger}\breve{g}_{\perp}(z_1)\breve{\vb*{J}}_{\perp}\breve{g}_{\perp}(z_1)\breve{g}_{\perp}(z_2)\breve{\eta}.
\end{align}\end{subequations}
This is consistent with the result obtained by transforming the density operator, which is defined as the elementary charge multiplied by the identity matrix in the $\mathbb{H}$ space, into the effective operator following the same procedure as shown in Fig.~\ref{fig:effsusdiag}(c).
Substituting these relations, the electric current operator is required to take the following form:
\begin{align}\begin{aligned}\label{eq:vertexe}
  &\breve{\vb*{\Gamma}}_{e} (\vb*{k}z_1,\vb*{k}z_2)\\
  &\quad= -\frac{e}{\hbar}\frac{\partial \breve{\mathcal{G}}^{-1}_{\eff}(\vb*{k},z_1)}{\partial \vb*{k}} + \left.(z_1-z_2)\frac{\partial\breve{\Gamma}_{e}^0(\vb*{k}'z_1,\vb*{k}z_2)}{\partial \vb*{k}'}\right|_{\vb*{k}'\rightarrow\vb*{k}}\\
  &\quad= \frac{e}{\hbar}\frac{\partial \breve{h}_{\eff}(\vb*{k},z_1)}{\partial \vb*{k}} + \left.(z_1-z_2)\frac{\partial\breve{\Gamma}_{e}^0(\vb*{k}'z_1,\vb*{k}z_2)}{\partial \vb*{k}'}\right|_{\vb*{k}'\rightarrow\vb*{k}}\\
  &\quad=
  \breve{\vb*{J}}_{\parallel}+ 
  \breve{\vb*{j}}^{\dagger}\breve{g}_{\perp}(z_2)\breve{\eta}+
  \breve{\eta}^{\dagger}\breve{g}_{\perp}(z_1)\breve{\vb*{j}}+\breve{\eta}^{\dagger}\breve{g}_{\perp}(z_1)\breve{\vb*{J}}_{\perp}\breve{g}_{\perp}(z_2)\breve{\eta},
\end{aligned}\end{align}
which indicates that the effective operator defined in Eq.~(\ref{eq:effECu}) satisfies the Ward--Takahashi identity.

Under the $\zeta$ approximation, We adopt the following symmetrized form, which preserves Hermiticity:
\begin{align}
  \breve{\Gamma}_{e}^0(\vb*{k}_1z_1,\vb*{k}_2z_2) &= e\frac{\breve{\zeta}^{-1}_{\vb*{k}_1}+\breve{\zeta}^{-1}_{\vb*{k}_2}}{2}.
\end{align}
Then,
\begin{subequations}\begin{align}
  \breve{\Gamma}_{e}^0(\vb*{k}z_1,\vb*{k}z_2) 
  &= e\breve{\zeta}^{-1},\\
  \left.\frac{\partial\breve{\Gamma}_{e}^0(\vb*{k}'z_1,\vb*{k}z_2)}{\partial \vb*{k}'}\right|_{\vb*{k}'\rightarrow\vb*{k}}
  &=\frac{e}{2}\frac{\partial \breve{\zeta}^{-1}}{\partial \vb*{k}}.
\end{align}\end{subequations}
Consequently, the electric current operator is evaluated as
\begin{align}\begin{aligned}\label{eq:vertexezeta}
  &\breve{\vb*{\Gamma}}_{e} (\vb*{k}z_1,\vb*{k}z_2)\\
  &\quad= -\frac{e}{\hbar}\frac{\partial \breve{\mathcal{G}}^{-1}_{\eff}(\vb*{k},z_1)}{\partial \vb*{k}} + \left.(z_1-z_2)\frac{\partial\breve{\Gamma}_{e}^0(\vb*{k}'z_1,\vb*{k}z_2)}{\partial \vb*{k}'}\right|_{\vb*{k}'\rightarrow\vb*{k}}\\
  &\quad= \frac{e}{\hbar}\frac{\partial \breve{h}_{\eff}(\vb*{k},0)}{\partial \vb*{k}}-z_1e\frac{\partial \breve{\zeta}^{-1}}{\partial \vb*{k}}+(z_1-z_2)\frac{e}{2}\frac{\partial \breve{\zeta}^{-1}}{\partial \vb*{k}}\\
  &\quad= e \left(\frac{1}{\hbar}\frac{\partial \breve{h}_{\eff}(\vb*{k},0)}{\partial \vb*{k}} -\frac{z_1+z_2}{2}\frac{\partial \breve{\zeta}^{-1}}{\partial \vb*{k}}\right),
\end{aligned}\end{align}
which demonstrates that the effective operator defined in Eq.~(\ref{eq:effOszeta}) satisfies the Ward--Takahashi identity.

\begin{figure*}[t]
  \centering
  \includegraphics[width=1.0\linewidth]{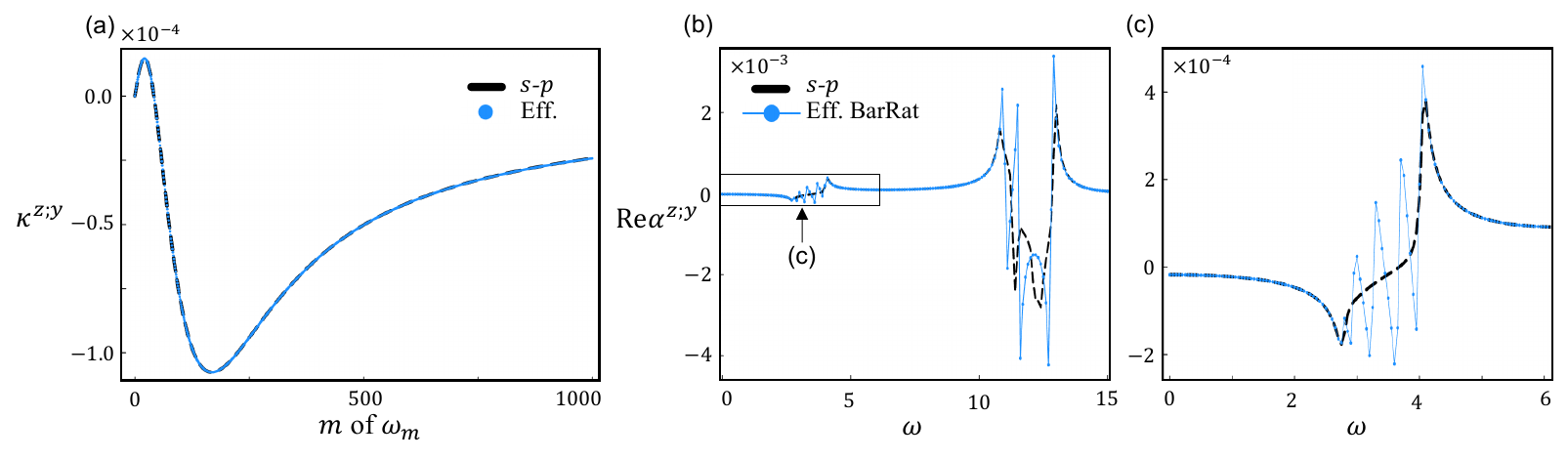}
  \caption{\label{fig:ACflow}
  (a) 
  The number of the bosonic Matsubara frequency ($m$ of $\omega_m$) dependence of the ME response function $\kappa^{z;y}(i\omega_m)$ in Eq.~(\ref{eq:effsus}).
  We also present the result calculated using the Kubo formula [Eq.~(\ref{eq:spKuboK})] in the $s$-$p$ model.
  The results are in agreement.
  [(b) and (c)]
  The $\omega$ dependence of the ME tensor in the effective model using the barycentric rational function approximation method.
  We also present the result using the Kubo formula in the $s$-$p$ model.
  While overall good agreement is observed, oscillations appear in the energy range associated with interband transitions.
  }
\end{figure*}

\paragraph*{Effective magnetoelectric operator.}

Finally, we consider the effective magnetoelectric operator $\hat{\Theta}^{\nu;\rho}$ with charge and spin-$z$ conservation.
The Ward--Takahashi identity for the second-order vertex function is given by
\begin{align}\begin{aligned}
  &\hbar k_{\varrho}' \, \breve{\Gamma}^{\varrho\varrho'}_{s_ze}(\vb*{k}_1z_1,\vb*{k}_2z_2;\vb*{k}'z')\\
  &= \breve{\Gamma}_{e}^{\varrho'}(\vb*{k}_1z_1,\vb*{k}_2-\vb*{k}'z_2-z')  \hat{s}_z
    - \hat{s}_z \breve{\Gamma}_{e}^{\varrho'}(\vb*{k}_1+\vb*{k}'z_1+z',\vb*{k}_2z_2),
\end{aligned}\end{align}
where $\varrho,\varrho'=0,x,y,z$, and the repeated index $\varrho$ is summed over with the metric signature $(1,-1,-1,-1)$.
When the output magnetization is along the $z$ direction, the effective magnetoelectric operator $\hat{\Theta}^{z;\rho}$ is related to the second-order vertex function as
\begin{align}\begin{aligned}
  &\hat{\Theta}^{z;\rho}(z;z')\\
  &\quad=  \frac{\muB \mathrm{g}}{\hbar} e^{i\vb*{k}\cdot\vb*{\hat{d}}} \breve{\Gamma}^{0\rho}_{s_ze}(\vb*{k}z,\vb*{k}z;\vb*{0}z') e^{-i\vb*{k}\cdot\vb*{\hat{d}}}
   +(\text{orbital term}).
\end{aligned}\end{align}
In the limit $\vb*{k}'\rightarrow\vb*{0}$, the Ward--Takahashi identity reduces to
\begin{align}\begin{aligned}
  &\hbar z' \, \breve{\Gamma}^{0\rho}_{s_ze}(\vb*{k}z,\vb*{k}z;\vb*{0}z')\\
  &\quad
  = \hat{s}_z \left( 
    \breve{\Gamma}_{e}^{\rho}(\vb*{k}z,\vb*{k}\,z-z') - \breve{\Gamma}_{e}^{\rho}(\vb*{k}\,z+z',\vb*{k}z) 
  \right).
\end{aligned}\end{align}
Using Eq.~(\ref{eq:vertexe}) and evaluating the right-hand side explicitly, we obtain
\begin{align}\begin{aligned}
  &\hat{s}_z \left( \breve{\Gamma}_{e}^{\rho}(\vb*{k}z,\vb*{k}\,z-z') - \breve{\Gamma}_{e}^{\rho}(\vb*{k}\,z+z',\vb*{k}z) \right)\\
  &\quad= \hbar z' \Big(
  \breve{j}^{\rho\dagger} \breve{g}_{\perp}(z-z') \hat{s}_z \breve{g}_{\perp}(z) \breve{\eta}
  + \breve{\eta}^{\dagger}\breve{g}_{\perp}(z) \hat{s}_z \breve{g}_{\perp}(z+z')\breve{j}^{\rho} \\
  &\qquad\qquad\quad+ \breve{\eta}^{\dagger} \breve{g}_{\perp}(z) \hat{s}_z \breve{g}_{\perp}(z+z') \breve{J}_{\perp}^{\rho} \breve{g}_{\perp}(z) \breve{\eta} \\
  &\qquad\qquad\quad+ \breve{\eta}^{\dagger} \breve{g}_{\perp}(z) \breve{J}_{\perp}^{\rho}  \breve{g}_{\perp}(z-z') \hat{s}_z \breve{g}_{\perp}(z) \breve{\eta}
  \Big),
\end{aligned}\end{align}
where we used $[\hat{P}_{\parallel},\hat{s}^z]=0$ and $[\hat{h},\hat{s}^z]=0$.
This agrees with the result obtained from the effective-operator transformation in Eq.~(\ref{eq:effThetaope}),

Under the $\zeta$ approximation, substituting Eq.~(\ref{eq:vertexezeta}) into the Ward--Takahashi identity yields
\begin{align}\begin{aligned}
  \breve{\Gamma}^{0\rho}_{s_ze}(\vb*{k}z,\vb*{k}z;\vb*{0}z')
  = \frac{e}{\hbar} \hat{s}_z \frac{\partial \breve{\zeta}^{-1}}{\partial \vb*{k}},
\end{aligned}\end{align}
which demonstrates that the eﬀective operator defined in Eq.~(\ref{eq:effOszeta}) satisfies the Ward--Takahashi identity.

\section{Results of the numerical analystic continuation} \label{sec:app:ACFlow}

In this Appendix, we present the results of numerical analytic continuation using the barycentric rational function approximation method~\cite{PhysRevB.111.125139}.
This method enables the reconstruction of real-frequency response functions from Matsubara-frequency data.

First, we compute the bosonic Matsubara frequency dependence of the ME response function $\kappa^{z;y}(i\omega_m)$ in the effective model based on Eq.~(\ref{eq:effsus}).
In the calculations, the fermionic Matsubara frequencies are summed over the range $n=-2000$ to $2000$, ensuring sufficient convergence.
For comparison, we also evaluate the ME response function in the full $s$--$p$ model using the Kubo formula:
\begin{align}\begin{aligned}\label{eq:spKuboK}
  \kappa^{z;y}(i\omega_m)
  =-\frac{1}{V}\sum_{k}\sum_{\alpha\beta}\frac{f(E_{k\alpha})-f(E_{k\beta})}{E_{k\alpha}-E_{k\beta}+i\hbar\omega_m}M^{z}_{k;\alpha\beta} J^{y}_{k;\beta\alpha}.
\end{aligned}\end{align}
As shown in Fig.~\ref{fig:ACflow}(a),
the results obtained from the effective model are in good agreement with those of the $s$--$p$ model, confirming the validity of the effective description at the Matsubara level.

Next, the Matsubara data are analytically continued to real frequencies using a fitting procedure based on the barycentric formula:
\begin{align}
  \kappa^{z;y}(z) \cong \left. \sum_{j} \frac{w_j f_j}{z - z_j} \middle/ \sum_{j} \frac{w_j}{z - z_j} \right. ,
\end{align}
where $\{z_j,f_j\}$ denote the sampled Matsubara data points and $w_j$ are the barycentric weights~\cite{PhysRevB.111.125139}.
In the numerical implementation, we use bosonic Matsubara frequencies in the range $m=0$ to $1000$.

Figures~\ref{fig:ACflow}(b) and (c) show the analytically continued results.
The effective model allows us to access a wide frequency range, including the high-energy region ($\omega \gg 6$) associated with the $p$-orbital sector.
However, spurious oscillations appear in certain frequency regions.  
These oscillations originate from the ill-posed nature of analytic continuation and the limited resolution of discrete Matsubara data.
Figure~\ref{fig:ACflow}(c) focuses on the low-energy $s$-orbital regime ($0 \leq \omega \leq 6$), where the physically relevant features are well captured.
The oscillatory behavior is most pronounced in the energy range between interband transition peaks. 
This can be attributed to the fact that the analytic continuation struggles to accurately reconstruct continuum-like spectral structures associated with interband processes from finite Matsubara data. 
Despite these limitations, the main features of the ME response, including peak positions and overall spectral shape, are reliably reproduced.

\nocite{apsrev42Control}
\bibliographystyle{apsrev4-2}
\bibliography{ZigzagME_bib.bib}

\end{document}